\documentclass[aps,10pt,twocolumn,showpacs,amsmath,amssymb,floatfix,superscriptaddress,longbibliography]{revtex4-1}
\usepackage{bm}
\usepackage{ulem}
\usepackage{xcolor}
\usepackage[colorlinks,linkcolor=blue,urlcolor=blue,citecolor=blue,anchorcolor=blue]{hyperref}
\usepackage[sort&compress]{natbib}
\usepackage{graphicx}
\usepackage{braket}
\usepackage{siunitx}
\usepackage{multirow}
\usepackage{siunitx}
\usepackage[utf8]{inputenc}

\newcommand{\tomas}[1]{\textcolor{black}{#1}}
\newcommand{\jj}[1]{\textcolor{black}{#1}}
\newcommand{\change}[1]{\textcolor{black}{#1}}
\newcommand{\changeByMing}[1]{\textcolor{black}{#1}}
\newcommand{\tomram}[1]{\textcolor{black}{#1}}

\begin{document}

\title{Scalable \tomas{multiphoton generation from cavity-synchronized} single-photon sources}

\author{Ming Li}
\affiliation{Instituto de F\'isica Fundamental, IFF-CSIC, Calle Serrano 113b, 28006 Madrid, Spain}
\affiliation{Department of Physics, Applied Optics Beijing Area Major Laboratory, Beijing Normal University, Beijing 100875, China}
\author{Juan Jos\'e Garc{\'i}a-Ripoll}
\affiliation{Instituto de F\'isica Fundamental, IFF-CSIC, Calle Serrano 113b, 28006 Madrid, Spain}
\author{Tom\'as Ramos}
\email{t.ramos.delrio@gmail.com}
\affiliation{Instituto de F\'isica Fundamental, IFF-CSIC, Calle Serrano 113b, 28006 Madrid, Spain}

\date{\today}

\begin{abstract}
\jj{We propose an efficient, scalable, and deterministic scheme to generate \changeByMing{multiple} indistinguishable photons over \changeByMing{independent} channels, on demand. Our design relies on multiple single-photon sources, each coupled to a waveguide, and all of them interact with a common cavity mode. The cavity synchronizes and triggers the simultaneous emission of one photon by each source, which are collected by the waveguides. For a state-of-the-art circuit QED implementation, this scheme supports the creation of single photons with purity, indistinguishability, and efficiency of $99\%$ at rates of $\sim $MHz. We also discuss conditions to \changeByMing{produce up to 100} photons simultaneously with \changeByMing{generation rates} of hundreds of kHz. This is orders of magnitude more efficient than previous demultiplexed sources for boson sampling and enables the realization of deterministic multi-photon sources and scalable quantum information processing with photons.}
\end{abstract}


\maketitle

\section{Introduction and main results}\label{sec:intro}

Efficient sources of single and indistinguishable photons \cite{eisaman_invited_2011,senellart_high-performance_2017,slussarenko_photonic_2019} are a fundamental requirement to perform all \jj{kinds} of quantum information tasks with photons: photonic quantum computation \cite{knill_scheme_2001,kok_linear_2007,tiecke_nanophotonic_2014,tiarks_optical_2016,takeda_toward_2019}, networking \cite{lodahl_quantum-dot_2017}, simulation \cite{aspuru-guzik_photonic_2012,hartmann_quantum_2016}, communication \cite{briegel_quantum_1998,sangouard_quantum_2011,llewellyn_chip--chip_2019}, cryptography \cite{gisin_quantum_2002}, metrology \cite{motes_linear_2015,ge_distributed_2018}, boson sampling \cite{brod_photonic_2019,wang_high-efficiency_2017,loredo_boson_2017,wang_boson_2019}, or even quantum optical neural networks \cite{steinbrecher_quantum_2019}. \jj{\tomas{Scaling up} these protocols require\tomas{s} the generation of a large product-state of $N\gg 1$ indistinguishable photons, 
\begin{align}
    \ket{\Psi_N}=\ket{1}_1\otimes\ket{1}_2\otimes\dots\otimes\ket{1}_N,\label{nphotonState}
\end{align}
propagating along $N$ or more channels. The generation of this state with high fidelity and efficiency \tomas{demands} the use of nearly deterministic, nearly identical, and perfectly synchronized single-photon sources (SPSs), each of them producing just one photon.}

\jj{The best on-demand SPSs for this task rely on few-level quantum systems\ \cite{fischer_dynamical_2016}, which can be deterministically excited\ \cite{he_theory_2006,he_theory_2006-1} and decay spontaneously, producing individual photons that are collected into the desired channels. The great experimental progress in controlling single quantum systems with cavity-enhanced light-matter interactions has allowed \tomas{many} realization\tomas{s of} nearly deterministic SPSs. A long list of setups includes single atoms \cite{mckeever_deterministic_2004,hijlkema_single-photon_2007}, single molecules \cite{brunel_triggered_1999,ahtee_molecules_2009,rezai_coherence_2018}, trapped ions \cite{almendros_bandwidth-tunable_2009,stute_tunable_2012,higginbottom_pure_2016,sosnova_trapped_2018}, and atomic ensembles \cite{chou_single-photon_2004,matsukevich_deterministic_2006,farrera_generation_2016,ripka_room-temperature_2018}, as well as solid-state systems  \cite{aharonovich_solid-state_2016} such as quantum dots \cite{santori_indistinguishable_2002,nawrath_coherence_2019,wang_-demand_2019,schnauber_indistinguishable_2019,thoma_exploring_2016,dusanowski_near-unity_2019,kirsanske_indistinguishable_2017,liu_high_2018,uppu_scalable_2020} or color centers in diamond \cite{rodiek_experimental_2017,sipahigil_indistinguishable_2014,alleaume_experimental_2004,gaebel_stable_2004,wu_room_2007,babinec_diamond_2010,tchernij_single-photon-emitting_2017,neu_single_2011,schroder_ultrabright_2011,wang_towards_2019}. In the  microwave regime, superconducting quantum circuits have been used to create nearly deterministic SPSs \cite{houck_generating_2007,pechal_microwave-controlled_2014,peng_tuneable_2016,forn-diaz_-demand_2017,gasparinetti_correlations_2017,pfaff_controlled_2017,eder_quantum_2018,zhou_tunable_2020}, which have the advantage of being externally tuneable \cite{peng_tuneable_2016,forn-diaz_-demand_2017,pfaff_controlled_2017}, fast \cite{pechal_microwave-controlled_2014,zhou_tunable_2020}, and readily integrated on-chip with very low losses \cite{zhou_tunable_2020}. Currently, quantum dots in micro-pillar cavities provide the best overall numbers including generation efficiencies, distinguishability, and single-photon purity \cite{reimer_quest_2019,wang_towards_2019,uppu_scalable_2020}, but it is hard to manufacture many of them identically, and tuning them also remains elusive for more than two emitters \cite{patel_tunable_2010,flagg_interference_2010,moczala-dusanowska_strain-tunable_2019,ellis_independent_2018,kambs_limitations_2018}.}

\jj{Despite remarkable experimental progress, scaling up to large number of identical SPSs remains a great challenge \cite{slussarenko_photonic_2019,wang_boson_2019,kaneda_high-efficiency_2019,hummel_efficient_2019,lenzini_active_2017}. Active optical {\it multiplexing} is a promising alternative which is based on repeating the single-photon generation---in time \cite{pittman_single_2002,kaneda_time-multiplexed_2015,kaneda_high-efficiency_2019,xiong_active_2016,mendoza_active_2016}, space \cite{migdall_tailoring_2002,ma_experimental_2011,shapiro_-demand_2007,collins_integrated_2013,francis-jones_all-fiber_2016,spring_chip-based_2017}, or frequency \cite{grimau_puigibert_heralded_2017,joshi_frequency_2018,hiemstra_pure_2019}---and then on synchronizing and rerouting the emitted photons using adaptive delay lines and switches. This method, originally developed to increase the efficiency of heralded SPSs\ \cite{pittman_single_2002,migdall_tailoring_2002,kaneda_high-efficiency_2019,christ_limits_2012}, has been adapted} to prepare $N$-photon states (\ref{nphotonState}) using just one nearly deterministic source\ \cite{lenzini_active_2017,wang_high-efficiency_2017,loredo_boson_2017,wang_boson_2019,anton_interfacing_2019,hummel_efficient_2019}. This {\it temporal-to-spatial demultiplexing} requires high quality \changeByMing{SPSs} that can emit a long stream \changeByMing{of} \jj{nearly indistinguishable photons\ \cite{ding_-demand_2016,somaschi_near-optimal_2016,liu_high_2018,dusanowski_near-unity_2019,uppu_scalable_2020}, as well as an accurate circuit that demultiplexes, routes and synchronizes the photons into multiple spatial channels. This technique enabled boson sampling with multi-photon states\ (\ref{nphotonState}) from  $N=3$ to $N=14$ single photons, albeit at a low photon rate of $\sim$kHz to $\sim$mHz \cite{lenzini_active_2017,wang_high-efficiency_2017,loredo_boson_2017,wang_boson_2019,anton_interfacing_2019,hummel_efficient_2019}.}

\jj{Apart from errors in the collection and detection of photons, \tomas{optical} multiplexing schemes are inherently \tomas{slow} and suffer from increased losses due to delay lines \cite{francis-jones_all-fiber_2016,nunn_enhancing_2013} and optical switches of the synchronizing circuit \cite{bonneau_effect_2015,lenzini_active_2017,gimeno-segovia_relative_2017}. \changeByMing{For a pulsed source with repetition rate $R$, the \tomas{$N$-photon} generation rate $C_N$ of producing the state $\ket{\psi_N}$ reads} \cite{lenzini_active_2017}
\begin{align}
\changeByMing{C_N= R} ({\cal P}_1)^N S_N,\label{scalingwithN2}
\end{align}
with \tomas{${\cal P}_1$ the probability of generating an individual photon \changeByMing{pulse} on each of the $N$ channels independently,} and $S_N$ the correlation error introduced by the synchronizing circuit. An ideal scheme should scale exponentially $(S_N=1)$, creating $N$ synchronous and completely uncorrelated photons. However, typical active demultiplexing schemes introduce errors that scale as} $S_N = \frac{1}{N}\left[ \eta^N+(N-1)\left(\frac{1-\eta}{N-1}\right)^N \right]$ \cite{lenzini_active_2017}, where $\eta$ characterizes the switching efficiency. Imperfect switching is a critical issue in any large photonic circuit \cite{li_resource_2015}, but even in the limit of lossless switches ($\eta=1$), the demultiplexing scheme introduces a \tomas{detrimental factor} $S_N=1/N$ to the scaling in Eq.~(\ref{scalingwithN}). This can significantly limit the achievable $N$-photon \changeByMing{generation rates} in practical applications \changeByMing{requiring $N\gg 1$} such as \tomram{single-photon} boson sampling.

\begin{figure}[t!]
\center
\includegraphics[width=0.9\columnwidth]{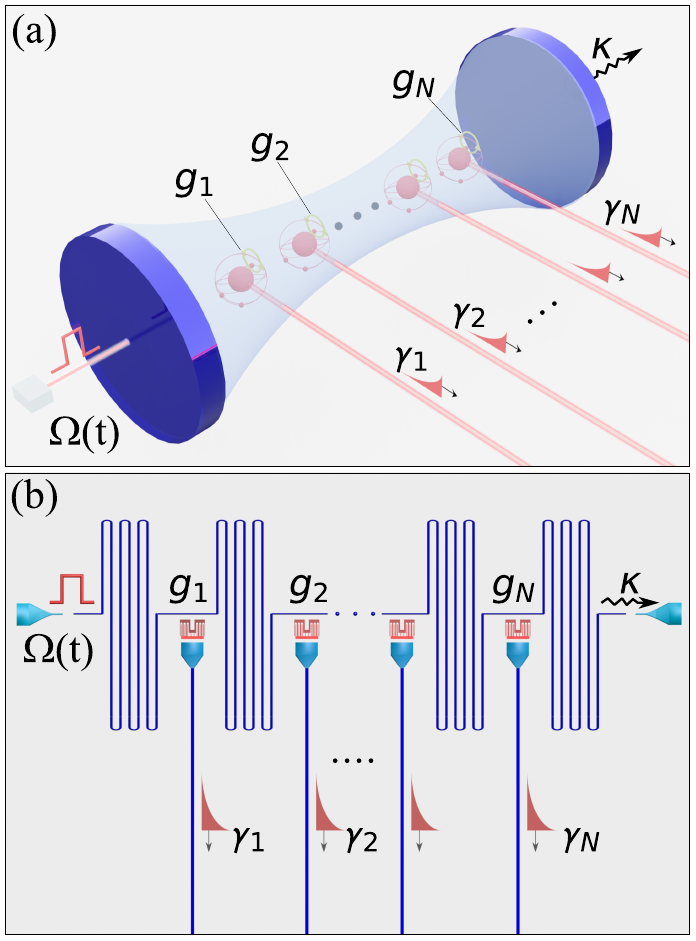}
\caption{Cavity-QED scheme for \tomas{deterministically generating a large number of synchronized and independent single-photons}. (a) Idealized implementation with $N$ two-level emitters coupled to a common driven cavity mode. Each emitter additionally couples to independent output channels, which collect the $N$ emitted single photons in parallel for later usage. (b) Efficient on-chip implementation using circuit-QED. Here, transmon qubits realize the $N$ two-level emitters, each of them capacitively coupled to an independent transmission-line antenna and to a common transmission-line resonator. Controlling the shape $\Omega(t)$ of the resonator drive, we can trigger all SPSs simultaneously and on demand. The $N$ emitted microwave single-photons are collected by the on-chip antennas with high efficiency. These antennas are also used to send a DC current to each qubit and calibrate their frequencies into resonance with the resonator drive.}
\label{fig:setup}
\end{figure}

In this work, we \jj{propose a scheme to synchronize deterministic SPSs and generate $N$ photons with \changeByMing{a nearly} negligible correlation error $S_N\approx 1$. The emitters reside in a \textit{bad} cavity, and interact both with the common electromagnetic mode and with independent waveguides [cf.~Fig.~\ref{fig:setup}]}. A strong \jj{coherent drive acting on the cavity excites the emitters in} a perfectly synchronized way. \jj{When the drive ends, all emitters relax, producing individual photons that are collected by the waveguides. If the emitters relax faster than the timescale of cavity-mediated correlations, the photons are nearly independent and approach the state $\ket{\psi_N}$. In a thorough study,} we identify the optimal parameter conditions to suppress residual cavity-mediated interactions and super-radiance. Using master equation and quantum trajectories simulations, we characterize the performance and scaling of the \jj{synchronized \tomas{multiphoton generation}, accounting for} imperfections and realistic noise sources. In particular, we find that the \changeByMing{probability $P_N$ of producing the $N$-photon state $\ket{\psi_N}$ in Eq.~(\ref{nphotonState})} scales nearly exponentially,
\begin{align}
\changeByMing{P_N= ({\cal P}_1)^N (1+D_N)},\label{scalingwithN}
\end{align}
\changeByMing{except for a small error $D_N=[P_N-({\cal P}_1)^N]/({\cal P}_1)^N\ll 1$, characterizing the residual correlation or dependence between photons. This small deviation is shown to} be quadratic in $N$:
\begin{align}
\changeByMing{D_N} = \epsilon N(N-1),\label{DNapprox}
\end{align}
\jj{with the correction factor $\epsilon$ stemming from residual cavity-induced correlations}. \changeByMing{We work out the parameter conditions to reach $\epsilon\sim 10^{-4}-10^{-7}$}, making the \tomas{multiphoton generation} scheme scalable up to $N\sim 100-1000\ll \epsilon^{-1/2}$, depending on the implementation and the noise sources. \changeByMing{The $N$-photon generation rate is \tomram{then obtained as} $C_N=RP_N$ with $R$ the repetition rate.}

To show the favorable scaling of \tomas{the} scheme in a realistic setup, we study a circuit-QED implementation with flux-tunable transmon emitters and microwave transmission lines\ \cite{dicarlo_demonstration_2009,barends_coherent_2013,pechal_microwave-controlled_2014}, \jj{considering} dephasing noise, internal loss, and disorder. As shown in Fig.~\ref{fig:setup}(b), this implementation is fully integrated on-chip, which allows the output antennas to have collection and transmission efficiencies above $99\%$ \cite{zhou_tunable_2020}. \changeByMing{Moreover, the use of the cavity as a common synchronizer has a low overhead compared to having and calibrating a large number of $N$ independent control lines and driving sources, which may lead to serious limitations in the wiring and the scalability of the superconducting chip \cite{tamate_toward_2022}. Using state-of-the-art parameters,} we show it is feasible to build microwave SPSs with single-photon purity and indistinguishability \changeByMing{of} $99\%$, as measured via standard Hanbury Brown and Twiss (HBT), and Hong-Ou-Mandel (HOM) experiments. Most importantly, we predict an overall single-photon probability or brightness of ${\cal P}_1\gtrsim 0.99$ \changeByMing{and the possibility to efficiently synchronize up to $N=100$ SPSs} with large multi-photon probabilities such as \changeByMing{$P_{10}\sim 0.90$}, $P_{30}\sim 0.72$, and $P_{100}\sim \changeByMing{0.16}$, \changeByMing{for $N=10$, $30$, and $100$ SPSs, respectively}. This means that for pulsed microwave SPSs with a repetition rate of $R\sim 0.4$ MHz, we can achieve $30$-photon generation rates of $C_{30}\sim \changeByMing{700}$ kHz, and $100$-photon rates of $C_{100}\sim \changeByMing{200}$ kHz. Remarkably, this is more than seven orders of magnitude more efficient than state-of-the-art boson sampling experiments with up to $N=14$ \tomram{single} photons\ \cite{wang_boson_2019}. The performance of our \jj{synchronized multiphoton generation} scheme benefits from the high efficiency of the superconducting circuit implementation, as well as from the absence of switches and other optical elements. We show that our predictions are robust to small inhomogeneities in the SPS parameters, \changeByMing{and to the multi-mode structure of the synchronizing cavity. Finally, we also consider the parameter conditions to synchronize up to $N=1000$ SPSs, which may be possible in more advanced circuit QED setups with weak and \tomram{highly} symmetric cavity-emitter couplings \tomram{\cite{huang_superconducting_2021, hazra_ring-resonator-based_2021}}.}

The paper is organized as follows. In Sec.~\ref{sec:model} we introduce the setup, model, main approximations, and identify the optimal parameter regime of the scheme. In Sec.~\ref{sec:performance}, we introduce realistic parameter sets for a circuit-QED implementation and quantify its performance via the \tomas{single- and multiphoton generation} efficiencies, single-photon purity, and indistinguishability. The most important results of the paper are presented in Sec.~\ref{sec:scaling}, where we characterize the scalability of the \tomas{synchronized $N$-photon generation}. In Sec.~\ref{sec:disorder} we analyze the effects of inhomogeneities and disorder in the SPS parameters \changeByMing{and in Sec.~\ref{sec:multi-mode} the effects of the multi-mode nature of the cavity.} Finally in Sec.~\ref{sec:conclusion} we summarize our conclusions. We complement the discussion with analytical and numerical methods to quantify the photon counting and photon correlations, which are presented in Appendices \ref{sec:effective_model}, \ref{sec:counting}, and \ref{sec:correlations}.

\section{Setup and synchronization}\label{sec:model}

In this section, we introduce a general cavity-QED model and possible nanophotonic implementations of the \tomas{multiphoton emitter} [cf.~Sec.~\ref{sec:setup}]. We then discuss how to achieve cavity-mediated synchronization [cf.~Sec.~\ref{sec:synchronization}], and finally we explain the basic working principle and the parameter conditions to achieve an efficient \tomas{multiphoton generation} [cf.~Sec.~\ref{sec:parameter_regime}].

\subsection{General cavity-QED model}\label{sec:setup}

\jj{Let us consider $N$ two-level systems or ``qubits'' coupled to an} optical or microwave cavity as shown in Figs.~\ref{fig:setup}(a) and (b). This cavity mode of frequency \tomas{$\omega_c$} is externally driven by a coherent field with time-dependent amplitude $\Omega(t)$ and frequency $\omega_d$. \change{The total Hamiltonian of the system reads
\begin{align}
H(t)={}& \omega_c a^{\dag}a +2\Omega(t)\cos(\omega_d t)(a+a^{\dag})\nonumber\\
{}&+ \frac{1}{2}\sum_{j=1}^N\omega_j^q\sigma_{j}^{z}+\sum_{j=1}^N g_j(a+a^{\dag})(\sigma_{j}^{-}+\sigma_{j}^{+}).
\label{SystemHamiltonian}
\end{align}}
Here, $a^{\dag}$, $a$ are the cavity creation and annihilation operators, whereas $\sigma_{j}^{+}=|e\rangle_j\langle g|$, $\sigma_{j}^{-}=|g\rangle_j\langle e|$, and $\sigma_j^z=|e\rangle_j\langle e|-|g\rangle_j\langle g|$ are standard Pauli operators for qubits $j=1,\dots, N$, with ground and excited states denoted by $|g\rangle_j$, and $|e\rangle_j$, respectively. \change{Each qubit $j$ has a possibly different frequency $\omega_q^j$ and a coupling $g_j$ to the common cavity mode. Note that we do not assume rotating wave approximation (RWA) to allow for a large cavity occupation $\langle a^\dag a\rangle\gg 1$.} We control the emitters by modulating the envelope of the cavity drive
\begin{align}
\Omega(t) = \Omega_0 f(t),
\end{align}
using a smooth square pulse $f(t)$ of duration $T$ and maximum amplitude $\Omega_0.$

Additionally, each qubit $j$ is coupled to an independent decay channel or ``antenna'', which collects the emitted photons [cf.~Fig.\ref{fig:setup}(a) and (b)]. We describe this qubit-antenna interaction in the Born-Markov approximation, \jj{introducing the $\gamma_j$ at which a qubit deposits photons into its antenna. We also introduce a photon loss rate $\gamma_{\rm loss}^j$ characterizing the emission of photons into any other unwanted channel. Finally, we consider the cavity decay rate $\kappa$ and introduce white noise dephasing rates $\gamma_{\phi}^j$ on each of the qubits. The complete dynamics of this open system are described with a master equation},
\begin{align}
\dot{\rho}(t)={}&-i[H(t),\rho]+\kappa{\cal{D}}[a]\rho+\sum_{j=1}^N \gamma_j{\cal{D}}[\sigma_{j}^{-}]\rho\nonumber\\\ 
{}&+\sum_{j=1}^N \gamma_{\rm loss}^j{\cal{D}}[\sigma_{j}^{-}]\rho+\sum_{j=1}^N 2\gamma_\phi^j{\cal{D}}[\sigma_{j}^{+}\sigma_{j}^{-}]\rho,\label{MasterEquation}
\end{align}
\jj{modelling the mixed quantum state of the qubits and the cavity mode $\rho(t)$, with the system Hamiltonian $H(t)$ from Eq.~(\ref{SystemHamiltonian}), and the Lindblad terms} ${\cal{D}}[x]\rho = x\rho x^{\dagger} - (x^{\dagger}x\rho + \rho x^{\dagger}x)/2$.

\jj{The idealized setup in Fig.~\ref{fig:setup}(a) admits various implementations. The two-level systems $\sigma_j^-$ could be neutral} atoms \cite{welte_photon-mediated_2018} or ions \cite{casabone_enhanced_2015} trapped inside an optical cavity field $a$ that is localized between macroscopic mirrors. In this prototypical cavity-QED implementation the individual decay channels of each qubit would require the use of high-aperture lenses \cite{araneda_interference_2018} or tapered nano-fibers \cite{reitz_coherence_2013} to collect the photons independently, which is very challenging to integrate and to scale-up with high collection efficiencies. Nanophotonic structures such as photonic crystals \cite{lodahl_interfacing_2015} or integrated photonic circuits \cite{anton_interfacing_2019,mehta_integrated_2020} are other promising platforms to realize \jj{our} setup. \jj{The common mode and the independent output channels can be integrated and scaled up. In this scenario, the main limitation arises from the creation of nearly identical or tuneable emitters such as quantum dots\ \cite{ellis_independent_2018}, or the trapping of many atoms \cite{goban_superradiance_2015} or ions \cite{mehta_integrated_2020} near these surfaces.}

\jj{However, in this work we will focus on circuit-QED to discuss an efficient and scalable integration of multiple two-level emitters with individual decay channels. As sketched in Fig.~\ref{fig:setup}(b), one may use superconducting transmon qubits as quantum emitters\ \cite{dicarlo_demonstration_2009,barends_coherent_2013,pechal_microwave-controlled_2014}. These qubits may be capacitively coupled to both} a common transmission-line resonator, as well as to individual transmission-line waveguides that efficiently collect the microwave single-photons \cite{zhou_tunable_2020}. In Secs.~\ref{sec:performance} and \ref{sec:scaling} we analyze in detail \changeByMing{the state-of-art parameters}, the performance and the scalability of this superconducting circuit platform.

\subsection{Cavity-mediated synchronization and residual correlations}\label{sec:synchronization}

The key mechanism to achieve an efficient cavity-mediated synchronization of the SPSs is to generate a large coherent state $|\alpha|\gg 1$ in the cavity mode. \change{This can be achieved via a strong drive, $|\Omega_0|\gg \kappa, |\Delta|$, where $\Delta = \omega_d-\omega_c$ is the cavity-drive detuning}. \jj{We model the resulting state using a displacement of the Fock operator
\begin{align}
    a = \alpha + \delta a,\label{cavityDispAlpha}
\end{align}
where the amplitude $\alpha(t)$ is given by the classical harmonic oscillator equation}
\begin{align}
\dot{\alpha}(t)={}&-\left[\kappa/2 + i\omega_c\right]\alpha(t)-2i\Omega_0f(t)\change{\cos(\omega_d t)}.\label{alphaEq}
\end{align}
Once the pulse is switched on ($f(t)=1$), $|\alpha(t)|$ grows and \change{stabilizes around} the steady state value  
\begin{align}
    |\alpha_{\rm ss}|= \frac{\Omega_0}{\sqrt{(\kappa/2)^2+\Delta^2}}\gg 1.\label{alphaSS}
\end{align}
\jj{Similarly, once we switch the drive off $(f(t\geq T)=0),$ the cavity displacement vanishes in a time scale $\sim 1/\kappa.$}

As shown in \jj{App.}~\ref{sec:displacement}, \jj{our control of the cavity displacement translates into a} cavity-mediated driving on all coupled qubits, given by the effective Hamiltonian, 
\begin{align}
\change{\Omega(t)\cos(\omega_d t)}(a+a^\dag)
    \rightarrow \sum_j g_j\change{{\rm Re}\lbrace \alpha(t)\rbrace}(\sigma_j^- + \sigma_j^+).\label{effDrivingcav}
\end{align}
\change{This is the mechanism that allows us to excite all SPSs simultaneously.} For our cavity-mediated control to succeed, the qubit back-action, cavity-induced interactions, and correlations between qubits must be suppressed. To do so, we restrict the couplings and detunings to keep the system in the bad-cavity limit
\begin{align}
    g_j,|\delta_j|\ll \kappa,|\Delta|\change{\ll \omega_j^q, \omega_c},\label{weakcouplingbadcavity}
\end{align}
\change{where $\delta_{j} = \omega_d - \omega_q^j$ is the detuning between qubit $j$ and cavity drive.} These conditions ensure that the cavity reaches a steady state~(\ref{alphaSS}), where small quantum fluctuations $\delta a\ll |\alpha_{\rm ss}|$ can be adiabatically eliminated \change{and fast oscillations neglected}. \change{As shown in App.~\ref{sec:adiabatic_elimination},} the dynamics of the qubit's reduced state $\tilde{\rho}(t)={\rm Tr}_c\lbrace \rho(t) \rbrace$ can then be modeled by an effective master equation, \change{which in the rotating frame with the drive frequency $\omega_d$ reads} 
\begin{align}
\dot{\tilde{\rho}}(t)={}&-i[\tilde{H}(t),\tilde{\rho}]+\sum_{j=1}^N \gamma_j{\cal{D}}[\sigma_{j}^{-}]\tilde{\rho}+\changeByMing{\sum_{j,l=1}^N\gamma^{jl}_{\rm cm}\tilde{{\cal D}}[\sigma_j^-,\sigma_l^-]\tilde{\rho}}\nonumber\\\ 
{}&+\sum_{j=1}^N \gamma_{\rm loss}^j{\cal{D}}[\sigma_{j}^{-}]\tilde{\rho}+\sum_{j=1}^N 2\gamma_\phi^j{\cal{D}}[\sigma_{j}^{+}\sigma_{j}^{-}]\tilde{\rho}.\label{master_D_AE}
\end{align}
\changeByMing{Here}, the effective Hamiltonian $\tilde{H}(t)$ has the form,
\begin{align}
\tilde{H}(t)={}& - \frac{1}{2}\sum_{j=1}^N(\delta_j-\delta^j_{\rm cm})\sigma_{j}^{z}+\sum_{j=1}^N \Omega_{\rm cm}^j f(t)(\sigma_{j}^{-}+\sigma_{j}^{+})\nonumber\\
{}&+\sum_{j>l} J^{jl}_{\rm cm}\left(\sigma_j^{+}\sigma_l^{-}+\sigma_l^{+}\sigma_j^{-}\right).
\label{H_D_AE}
\end{align}
\change{with $\Omega_{\rm cm}^j$ the cavity-mediated driving given by}
\begin{align}
    \Omega_{\rm cm}^j ={}& |\alpha_{\rm ss}|g_j,\label{cm_Omega}
\end{align}
with the steady state coherent amplitude $|\alpha_{\rm ss}|$ given in Eq.~(\ref{alphaSS}). In addition, the qubits \jj{experience} a Lamb-shift
\begin{align}
    \delta^j_{\rm cm} ={}& \frac{(g_j)^{2}\Delta}{ (\kappa/2)^2+\Delta^2},\label{cm_delta}
\end{align}
and undergo long-range cavity-mediated interactions, with \tomas{couplings}
\begin{align}
J^{jl}_{\rm cm} ={}& \frac{g_j g_l \Delta}{(\kappa/2)^2+\Delta^2}.\label{cm_J}
\end{align}
\jj{The cavity also} induces a collective decay or ``superradiance'' on the qubits, described by a generalized Lindblad term \changeByMing{$\tilde{{\cal D}}[x,y]\rho=x\rho y^\dag -(y^\dag x\rho+\rho y^\dag x)/2$ in Eq.~(\ref{master_D_AE})} and \jj{cavity-mediated decay rates}
\begin{align}
\changeByMing{\gamma^{jl}_{\rm cm}} ={}& \frac{\changeByMing{g_jg_l}\kappa}{ (\kappa/2)^2+\Delta^2}.\label{cm_gamma}
\end{align}
\jj{Finally, the effective dynamics of the qubits also include local decay and dephasing rates $\gamma_j$, $\gamma_j^{\rm loss}$, and $\gamma_\phi$, as described by the normal Lindblad terms in Eq.~(\ref{master_D_AE}).}

\subsection{Synchronization dynamics and parameter regime}\label{sec:parameter_regime}

\jj{Our goal is to realize the synchronized excitation and emission of each qubit so that they act as nearly independent SPSs. The first stage of operation involves exciting each two-level system from the ground state $|g\rangle_j$ to the excited state $|e\rangle_j$ via a fast cavity-mediated $\pi$-pulse of duration $T\sim \pi/(2\Omega_{\rm cm}^j)$. We then expect that all emitters will produce synchronized photons, one on each of the antennas, on a timescale $t\gg 1/\gamma_j$. To ensure that this \tomas{procedure efficiently}} generates $N$ nearly indistinguishable and independent single-photons, the system parameters must satisfy the conditions,
\begin{align}
\Omega_{\rm cm} \gg{}& \gamma\gg J_{\rm cm},\gamma_{\rm cm},\gamma_{\rm loss},\gamma_{\phi},\label{parameter_conditions}
\end{align}
and $\delta = \delta_{\rm cm}$, as explained in the following. 

First, to achieve synchronization and indistinguishability of the photon emissions, the system parameters should be as homogeneous as possible, especially the qubit frequencies $\omega_q\approx \omega_q^j$, couplings $g\approx g_j$, and antenna decays $\gamma\approx \gamma_j$. In the remainder of the paper, we will thus assume that all system parameters are homogeneous, except in Sec.~\ref{sec:disorder}, where we analyze inhomogeneities and disorder.

Second, to achieve high efficiency and purity of single-photon emission we need to drive the emitters on resonance ($\delta=\delta_{\rm cm}$). \jj{We must also excite the qubits very fast ($\Omega_{\rm cm}\gg \gamma$) to suppress events where the qubit emits two photons---it creates a photon during the excitation pulse, gets excited by the remaining of the pulse, and emits a second photon---. Note that} we can mitigate this effect using three-level emitters as SPSs \cite{fischer_dynamical_2016}, but this is not the focus of the present work. 

Third, to achieve nearly independent \jj{photons in a product state (\ref{nphotonState}), the photons must be created faster than the speed of the cavity mediated correlations and interactions} ($\gamma\gg J_{\rm cm},\gamma_{\rm cm}$). This is the only intrinsic limitation of the proposed \tomas{multiphoton} scheme, \change{and its effect can be mitigated in the case of a bad cavity $g\ll \kappa$ with strong coherent drive $|\alpha_{\rm ss}|\gg 1$.} In Sec.~\ref{sec:performance}-\ref{sec:scaling} we show that \jj{state-of-the-art circuit-QED setups satisfy this requirement and allow} the synchronization of $N\sim 100-1000$ nearly independent SPSs. 

The fourth condition, $\gamma\gg \gamma_{\rm loss}$, \jj{ensures an efficient collection of photons by the antennas, with minimal losses}, while the last condition, $\gamma\gg\gamma_\phi$, \jj{ensures that those photons are phase-coherent and indistinguishable} as detected by HOM interference [cf.~Sec.~\ref{sec:indistinguishability}]. \jj{The degree to which we satisfy these conditions is limited by the technology of the SPSs---atoms, dots, superconductors, etc---, but we will see that they are well met by} state-of-the-art circuit-QED setups [cf.~Sec.~\ref{sec:performance}].

\begin{figure}[t!]
\includegraphics[width=1\columnwidth]{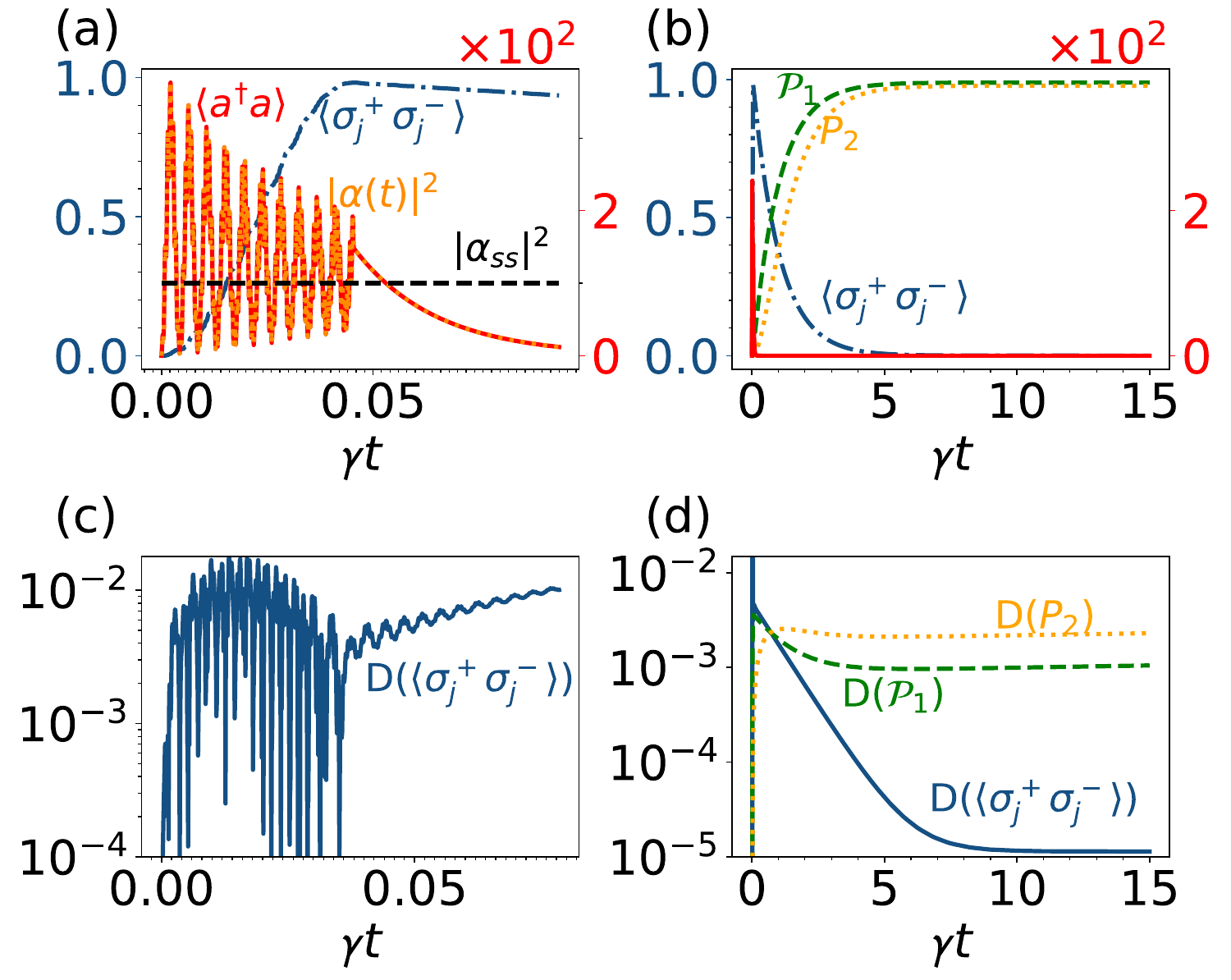}
\caption{Synchronization of two deterministic SPSs via a cavity. (a) Short time dynamics: The cavity pulse is switched on $(f(t)=1)$ and a large coherent state with $\langle a^\dag a \rangle\sim |\alpha_{\rm ss}|^2\sim \changeByMing{10^2}$ photons is quickly created in the cavity \change{with fast oscillations due \changeByMing{to the large detuning $\Delta$ and} the non-RWA terms [cf.~red solid and black dashed lines]}. This induces an effective cavity-mediated driving $\Omega_{\rm cm}=|\alpha_{\rm ss}|g$, which resonantly excites both qubits $\langle \sigma_j^+\sigma_j^-\rangle$ synchronously [cf.~blue dash-dotted line]. After a $\pi$-pulse time $T\sim\pi/(2\Omega_{\rm cm})$, the cavity pulse is switched off $(f(t)=0)$, $\langle a^\dag a \rangle$ decays in a fast timescale $\sim 1/\kappa$, and both qubits are left in their excited states $\langle \sigma_j^+\sigma_j^-\rangle\approx 1$. (b) Long-time dynamics: Both qubits spontaneously decay in a timescale $t\sim 15/\gamma$, and each of them generates a single-photon with \tomas{probability} ${\cal P}_1\approx 0.99$ on its antenna [cf.~ green dashed line]. Since these spontaneous emission processes are much faster than the timescale for the cavity-mediated interactions ($J_{\rm cm}/\gamma\sim \changeByMing{10^{-2}}$) and collective decay ($\gamma_{\rm cm}/\gamma\sim 10^{-4}$), both single-photons are nearly perfectly synchronized and independent as manifested by the two-photon probability satisfying $P_2\approx ({\cal P}_1)^2\approx 0.98$ [cf.~yellow dotted line]. \change{(c)-(d) Deviations $D(x)=|x_{\rm exact}-x_{\rm eff}|$ between the exact (\ref{MasterEquation}) and effective (\ref{master_D_AE}) models for the qubit occupations $x=\langle\sigma_j^+\sigma_j^-\rangle$, and photon generation probabilities $x={\cal P}_1,P_2$ as a function of time. We observe great agreement and no heating effects despite the large cavity driving strength $\Omega_0=4\omega_d \gg \kappa,|\Delta|$. These and the rest of} the parameters correspond to a state-of-the-art circuit-QED implementation shown in row \changeByMing{A} of Table~\ref{parameters} \change{and discussed in Sec.~\ref{sec:parameter_sets}}.}
\label{fig:dynamics}
\end{figure}


\begin{table*}[t]
\center
\begin{tabular}{|c|c|c|c|c||c|c|c||c|c|c||c|c|c|c|}
\hline\hline
   & $g/2\pi$ & $\kappa/2\pi$ & $\Delta/2\pi$ & $|\alpha_{\rm ss}|^2$ & $\gamma/2\pi$& $\gamma_{\rm loss}/2\pi$  & $\gamma_\phi/2\pi$ & $\gamma/\Omega_{\rm cm}$ &
  $\gamma_{\rm cm}/\gamma$ &
  $J_{\rm cm}/\gamma$ &
   ${\cal P}_1$ & $P_{2}$& $g^{(2)}_{\rm HBT}[0]$ & $g^{(2)}_{\rm HOM}[0]$\\

 & $[{\rm MHz}]$ & $[{\rm MHz}]$ & $[{\rm MHz}]$ & & $[{\rm kHz}]$ & $[{\rm kHz}]$ & $[{\rm kHz}]$ & & & & & & & \\
\hline
A & \changeByMing{4.0} & \changeByMing{50} & \changeByMing{1400} &\changeByMing{$10^{2}$} &\changeByMing{966} &\changeByMing{5} &\changeByMing{10} & \changeByMing{$2.4\cdot 10^{-2}$} & \changeByMing{$4.2\cdot 10^{-4}$} & \changeByMing{$1.2\cdot 10^{-2}$}& \changeByMing{0.989}&\changeByMing{0.979} &\changeByMing{$1.4\cdot 10^{-2}$} & \changeByMing{$1.4\cdot 10^{-2}$}\\
\hline
B & \changeByMing{4.0} & \changeByMing{50} & \changeByMing{1400} & \changeByMing{$10^{2}$} &\changeByMing{3000} &\changeByMing{10} &\changeByMing{20} & \changeByMing{$7.5\cdot 10^{-2}$} & \changeByMing{$1.4\cdot 10^{-4}$} & \changeByMing{$3.8 \cdot 10^{-3}$}& \changeByMing{0.982}& \changeByMing{0.964}& \changeByMing{$2.8 \cdot 10^{-2}$} & \changeByMing{$7.0 \cdot 10^{-3}$}\\
\hline
\hline
C & 0.2 & 400 & 50 & $ 4\cdot 10^4$ &400 &1 &1 &$1.0\cdot 10^{-2}$ & $9.4\cdot 10^{-4}$&$1.2\cdot 10^{-4}$ &0.994 &0.989 & $4.4\cdot 10^{-3}$& $2.5 \cdot 10^{-3}$\\
\hline
D & 0.2 & 400 & 50 & $4\cdot 10^4$ &1000 &1 &1 &$2.6\cdot 10^{-2}$ & $3.8\cdot 10^{-4}$&$4.7\cdot 10^{-5}$ & 0.991& 0.983& $1.3\cdot 10^{-2}$& $7.7\cdot 10^{-3}$\\

\hline\hline
\end{tabular}
\caption{Parameters and figures of merit for a circuit-QED implementation of two synchronized SPSs. We consider four parameter sets: \changeByMing{A and B correspond state-of-the-art parameters $g,\kappa,\Delta$ and $|\alpha_{\rm ss}|^2$ for different values of antenna couplings $\gamma$ and decoherence rates $\gamma_{\rm loss}$, $\gamma_\phi$. C and D correspond to more challenging long-term parameters due to the lower coupling $g$ and larger $|\alpha_{\rm ss}|^2$. All parameters lead to the same effective cavity-mediated drive $\Omega_{\rm cm}/2\pi\approx 40$ MHz.} The detuning of the qubits is chosen to compensate for \changeByMing{any cavity-mediated shift, including the Lamb-shift $\delta_q\sim\delta_{\rm cm}$ or residual Bloch-Siegert shifts}. Depending on the \changeByMing{couplings and} decoherence rates of sets A-D, the inequalities in Eq.~(\ref{parameter_conditions}) are satisfied differently, as shown in columns \changeByMing{8th to 10th}, and this influences the performance of figures of merit \changeByMing{in the last four columns}: single-photon generation probability ${\cal P}_1$ [cf.~Sec.~\ref{sec:efficiency_SPS}], synchronized two-photon generation efficiency or probability $P_2\approx ({\cal P}_1)^2$ [cf.~Sec.~\ref{sec:efficiency_NP}], multi-photon contamination $g^{(2)}_{\rm HBT}[0]$ [cf.~Sec.~\ref{sec:purity}], and photon distinguishability $g^{(2)}_{\rm HOM}[0]$ [cf.~Sec.~\ref{sec:indistinguishability}]. All these quantities are calculated using the full master equation (\ref{MasterEquation}), the methods in Appendices \ref{sec:counting_ME} and \ref{sec:correlations}, and evolving up to a final time $t = 15/ \gamma$.
\label{parameters}}
\end{table*}

\jj{Fig.~\ref{fig:dynamics}(a)-(b) illustrates the two processes in the cavity-mediated synchronization method, using two synchronized SPSs with parameters that satisfy Eq.\ (\ref{parameter_conditions}).} Fig.~\ref{fig:dynamics}(a) displays the short time dynamics: the creation of a large coherent state in the cavity \changeByMing{which induces an oscillatory occupation around the large steady-state value}, $\langle a^\dag a\rangle\sim |\alpha_{\rm ss}|^2\sim \changeByMing{10^2}$ [cf.~red solid line \changeByMing{and dashed black line], while mediating} the \jj{simultaneous excitation of the} qubits \changeByMing{up to} $\langle \sigma_j^+\sigma_j^-\rangle\approx 1$ \changeByMing{(when the cavity driving is switched off)} [cf.~blue dash-dotted line]. Fig.~\ref{fig:dynamics}(b) \jj{shows the long-time dynamics, in which the photons are generated.} \tomas{Each emitter decays almost independently in a timescale $t\sim 15/\gamma$, depositing a photon into its own antenna [cf.~blue dash-dotted line]. To quantify the efficiency of these emission processes, we display the single-photon and synchronized two-photon generation probabilities, ${\cal P}_1$ and $P_2$, calculated using the definitions and methods in Secs.~\ref{sec:efficiency_SPS}-\ref{sec:efficiency_NP}. For non-optimized parameters satisfying Eq.~(\ref{parameter_conditions}), we reach ${\cal P}_1\approx 0.99$ [cf.~green dashed line] and $P_2\approx ({\cal P}_1)^2\approx 0.98$ [cf.~yellow dotted line]. This demonstrates that both emitted photons are highly independent of each other [$S_2\approx 1$ in Eq.~(\ref{scalingwithN})] and that cavity-induced correlations are indeed negligible during photon emission processes ($J_{\rm cm}/\gamma\sim \changeByMing{10^{-2}}$ and $\gamma_{\rm cm}/\gamma\sim 10^{-4}$).} 

\change{We performed the numerical simulations in Figs.~\ref{fig:dynamics}(a)-(b) using both the full model in Eqs.~(\ref{SystemHamiltonian})-(\ref{MasterEquation}) and the effective model in Eqs.~(\ref{master_D_AE})-(\ref{cm_gamma}). They show an excellent agreement with small deviations $D(x)=|x_{\rm exact}-x_{\rm eff}|<10^{-2}$ in the qubit and photon counting observables [cf.~Fig.~\ref{fig:dynamics}(c)-(d)], despite the large cavity population of $\langle a^\dag a\rangle\sim \changeByMing{10^2}\gg 1$. This demonstrates that the strong cavity driving $\Omega_0\gg \kappa,|\Delta|$ does not introduce heating in the system, but only creates a large coherent state in the cavity with some fast oscillations around the mean value $|\alpha_{\rm ss}|^2\gg 1$ [cf.~solid red and dashed black lines in Fig.~\ref{fig:dynamics}(a)]. Since these fast oscillations have no detrimental impact on the functionality of our multiphoton generation scheme, in the remainder of the paper we use the simpler effective model (\ref{master_D_AE}) which uses the mean steady value (\ref{alphaSS}) for describing the effect of the cavity field (see App.~\ref{sec:effective_model} for more details on the approximations involved in the model).} 

\section{Performance of a circuit-QED implementation}\label{sec:performance}

In this section, we describe typical state-of-the-art parameters to realize the \tomas{synchronized multiphoton emitter} in a circuit-QED implementation [cf.~Sec.~\ref{sec:parameter_sets}]. Subsequently, we quantify the performance of the scheme using four figures of merit: (i) efficiency of single-photon generation ${\cal P}_1$ [cf.~Sec.~\ref{sec:efficiency_SPS}], (ii) efficiency of synchronized $N$-photon generation $P_N$ [cf.~Sec.~\ref{sec:efficiency_NP}], (iii) single-photon purity as quantified by HBT correlations, $p=1-g^{(2)}_{\rm HBT}[0]$ [cf.~Sec.~\ref{sec:purity}], and (iv) photon indistinguishability as quantified by HOM interference, $I=1-g^{(2)}_{\rm HOM}[0]$ [cf.~Sec.~\ref{sec:indistinguishability}].

\subsection{State-of-the-art parameters}\label{sec:parameter_sets}

In Fig.~\ref{fig:setup}(b) we \jj{sketch a possible setup \tomas{to implement the synchronized multiphoton emitter} using superconducting circuits.} For the cavity mode $a$, we consider a superconducting transmission line resonator with decay rate $\kappa/2\pi = \changeByMing{50} {\rm MHz}$, and resonance frequency $\omega_c$ on the order of a few GHz. These superconducting resonators \jj{support strong drives and can hold} a large number of photons in steady state \change{$|\alpha_{\rm ss}|^2\gg 1$ \cite{pietikainen_observation_2017,jenkins_nanoscale_2014,macri_spin_2020}, as it is required in our scheme}. \changeByMing{In particular, we consider $|\alpha_{\rm ss}|^2\approx 100$ \tomram{of average} photons in steady state, induced by a cavity driving of $\Omega_0/2\pi = 14 {\rm GHz}$ and a detuning of $\Delta/2\pi=  1400 {\rm MHz}$. For standard transmission line resonators such as in Ref.~\cite{pietikainen_observation_2017}, this order of driving strength $\Omega_0$ is obtained with a voltage source of power $P_{\rm in}\sim -50$ dBm.}

As two-level emitters, we consider flux-tunable transmon qubits\ \cite{dicarlo_demonstration_2009,barends_coherent_2013,pechal_microwave-controlled_2014,wang_controllable_2020} weakly coupled to the resonator mode with strength $g/2\pi = \changeByMing{4} {\rm MHz}$. \jj{The cavity mediated driving amplitude} \tomram{then reads} $\Omega_{\rm cm}/2\pi=|\alpha_{\rm ss}|g/2\pi\approx 40$MHz. This is strong but still \jj{smaller} than typical transmon anharmonicities $U/2\pi \sim 300-400 {\rm MHz}$ \cite{pechal_microwave-controlled_2014,mckay_three-qubit_2019,zhou_tunable_2020}, thus preventing leakage to higher excitation states. The frequency of the qubits---which will be in the few GHz range---can be fine-tuned by sending a DC current through the antenna that is coupled to each qubit [cf.~Fig.~\ref{fig:setup}(b)]. This is required to bring the qubits in resonance with the external drive, and to compensate \changeByMing{any cavity-induced shifts such as Lamb-shifts $\delta_q\sim\delta_{\rm cm}$ or residual Bloch-Siegert shifts \cite{pietikainen_observation_2017}} [See Sec.~\ref{sec:disorder} for small deviations of this condition].

The antenna of each qubit is not only used for frequency calibration, but its main purpose is to collect the \jj{photons} emitted by each qubit [cf.~Fig.~\ref{fig:setup}(b)]. \jj{The coupling rate $\gamma$ between the qubit and antenna} can be chosen to \changeByMing{optimally fulfill the inequalities in Eq.~(\ref{parameter_conditions}) and it also determines the time-scale of photon emission $t\sim 15/\gamma$. For the circuit QED parameters discussed above, we consider two values of $\gamma$: Parameter set A with $\gamma/2\pi\approx 0.97$ MHz optimized to maximize the emission probabilities ${\cal P}_1\approx 0.99$ and $P_2\approx 0.98$, as well as parameter set B with $\gamma/2\pi=3$ MHz optimized for \tomram{faster emission rate $C_2\approx 1.2$ MHz} and lower cavity-mediated effects, $J_{\rm cm}/\gamma\sim 10^{-3}$. The value of $\gamma$ also limits the maximum unwanted decoherence rates $\gamma_{\rm loss}$ and $\gamma_\phi$ that can be tolerated by the device to work properly. In particular, we consider $\gamma_{\rm loss},\gamma_{\phi}\sim 2\pi \cdot 10 {\rm kHz} \ll \gamma$ as realized in state-of-art experiments \cite{barends_coherent_2013,caldwell_parametrically_2018,mckay_three-qubit_2019,andersen_repeated_2019}.}

\changeByMing{Table \ref{parameters} summarizes all the parameter sets considered in this work and their main figures of merit as discussed below. Note that we also consider extra parameter sets C and D, which reduce cavity-mediated effects by two orders of magnitude compared to A and B, respectively. This is achieved by having a 20-fold lower cavity-emitter coupling rate $g/2\pi=0.2$ MHz compensated by a larger cavity mean photon number $|\alpha_{\rm ss}|^2\sim 4\cdot 10^4$ so that the effective drive on the qubits is preserved, $\Omega_{\rm cm}=g|\alpha_{\rm ss}|\sim 2\pi\cdot 40$MHz. Engineering such a low coupling with high homogeneity on all emitters is challenging with current circuit QED technology \tomram{because the coupling becomes} of the same order as parasitic capacitive couplings typically found in superconducting circuits \cite{wang_controllable_2020}. Nevertheless, reducing the inhomogeneity of these parasitic couplings may be possible in the long term via highly symmetric ring resonators \cite{huang_superconducting_2021,hazra_ring-resonator-based_2021}. Parameters C and D consider this possibility and below we discuss how the scalablity can reach up to $N\sim 1000$ SPSs in this case. The second important requirement of parameters C and D is the large mean photon displacement $|\alpha_{\rm ss}|^2\sim 4\cdot 10^4$, but this is well below the values $10^5-10^9$ photons reported with current technology \cite{pietikainen_observation_2017,jenkins_nanoscale_2014} and is achieved by having a lower detuning $\Delta/2\pi = 50$ MHz, larger cavity drive $\Omega_0/2\pi = 40$ GHz and \tomram{a decay of} $\kappa/2\pi=400$ MHz. We note the strong drive required by our scheme is not harmful to the superconducting device since most of its energy is reflected by the cavity and it mainly induces a coherent displacement with no thermal component \tomram{(see App.~\ref{sec:effective_model} for more details)}. In addition, in Sec.~\ref{sec:LongCavity}, we show that using a long superconducting resonator $L\sim 66$cm \cite{sundaresan_beyond_2015,chang_remote_2020} the multi-mode structure of the cavity allows to reduce the required external drive down to $\Omega_0/2\pi\sim 2.6 $ GHz while having the same effective quantities considered in sets C and D}.

In the remainder of the paper, we use the four parameter sets A-D in Table \ref{parameters} to characterize the performance and scaling of the synchronized SPSs for this circuit-QED implementation. 

\begin{figure*}[t]
\includegraphics[width=0.95\textwidth]{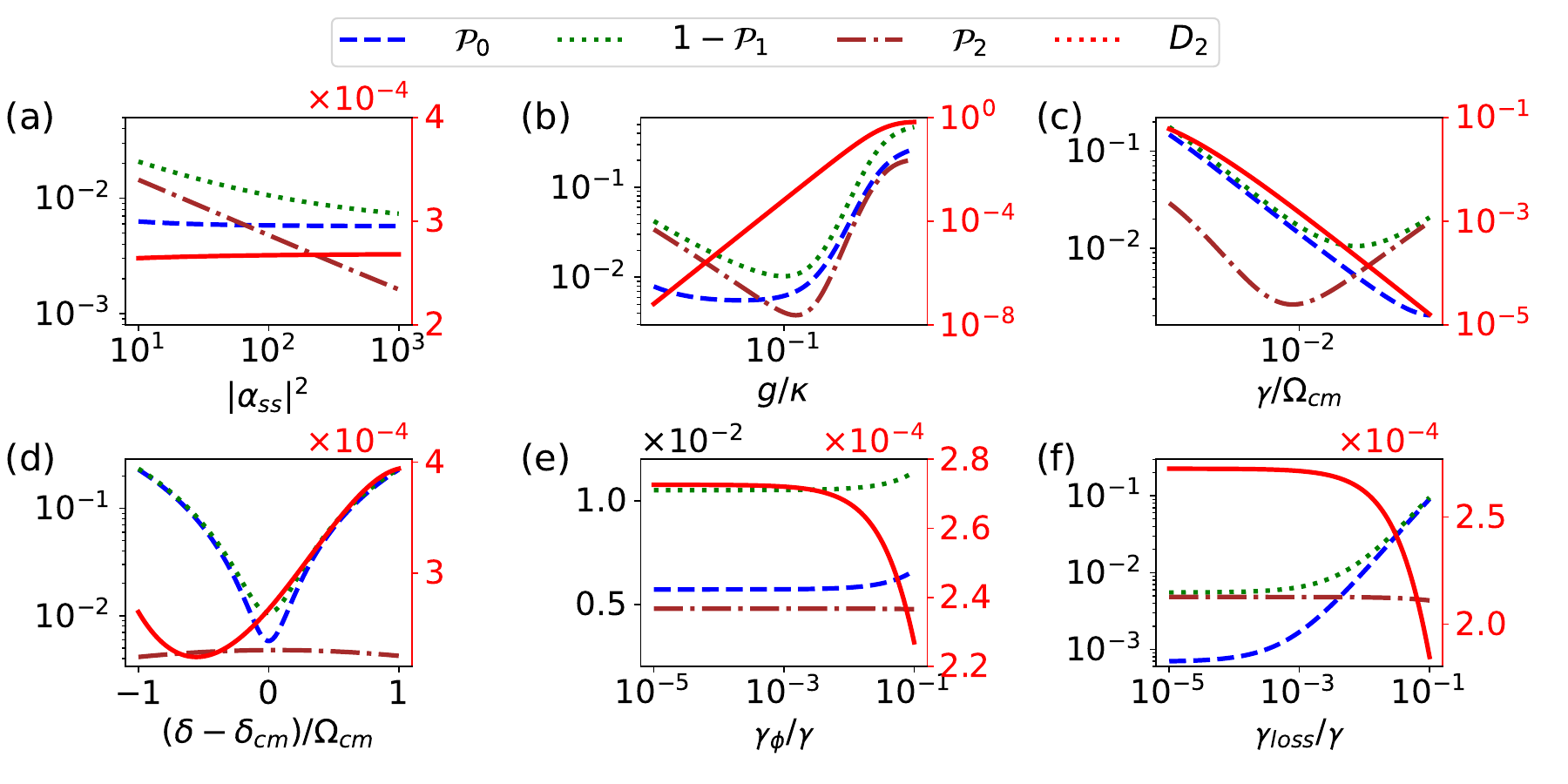}
\caption{Few-photon statistics ${\cal P}_n$ and two-photon \tomas{dependence} $D_2$ as a function of various system parameters such as (a) cavity occupation $|\alpha_{\rm ss}|^2$, (b) qubit-cavity coupling $g/\kappa$, (c) qubit-antenna decay $\gamma/\Omega_{\rm cm}$, (d) qubit-drive detuning $(\delta-\delta_{\rm cm})/\Omega_{\rm cm}$, (e) dephasing $\gamma_{\phi}/\gamma$, and (f) relaxation loss $\gamma_{\rm loss}/\gamma$. The probabilities ${\cal P}_0$ (green), $1-{\cal P}_1$ (blue), and ${\cal P}_2$ (brown) are plotted using the left vertical axis, whereas the two-photon \tomas{dependence} $D_2$ (red), corresponds the right vertical axis. \changeByMing{All quantities are computed with} the effective master equation (\ref{master_D_AE}) using parameter set \changeByMing{A} of Table~\ref{parameters}, except for the quantity that is changed on each subplot.}
\label{fig:conditions}
\end{figure*}

\subsection{Efficiency of single-photon generation}\label{sec:efficiency_SPS}

\jj{Without loss of generality, in our discussion about photon generation efficiencies or probabilities, we assume perfect photo-detectors\ \footnote{Detector inefficiencies $\eta_{D}<1$ give rise to an extra exponential factor $\sim(\eta_{D})^N$ \cite{lenzini_active_2017} in the $N$-photon probability (\ref{scalingwithN}). See also Refs.~\cite{eisaman_invited_2011,slussarenko_photonic_2019} for more details.}. In this scenario, the efficiency of a single photon source can be extracted from the photon emission statistics. For a single emitter, ${\cal P}_1$ is the probability of generating exactly an isolated photon in a Fock state\ \cite{senellart_high-performance_2017}. Similarly, we define ${\cal P}_n^j$ as the probability for the synchronized SPS to emit a Fock state of $n$ photons in the $j$-th antenna. Formally,} 
\begin{align}
    {\cal P}_n^j = {\rm Tr}\lbrace \Lambda_{n}^j \rho_{\rm ext}(t)\rbrace,\label{efficiencyDef}
\end{align}
where $\rho_{\rm ext}(t)$ is the state of the extended system including the photons emitted in all antennas. The operator $\Lambda_n^j = |n\rangle_j\langle n|$ projects $\rho_{\rm ext}(t)$ on the Fock state $|n\rangle_j$ of $n$ photons propagating in the independent antenna $j$. \jj{Computing  ${\cal P}_1$ or ${\cal P}_n^j$ seems to require a large Hilbert space, but there are shortcuts, such as the quantum jump formalism or our new photon counting approach based on master equations [cf.~App.~\ref{sec:counting}].}

A \tomas{nearly deterministic} single-photon source should have ${\cal P}_1\approx 1$, other few-photon probabilities ${\cal P}_0$ and ${\cal P}_{2}$ should be strongly suppressed, and ${\cal P}_{n\geq 3}\approx 0$ negligible. \jj{This also applies to each of the probabilities ${\cal P}_n^j$ in our synchronized setup, if we want it to operate as a collection of $N$ independent emitters.} To verify this, in Figs.~\ref{fig:conditions}(a)-(f) we have computed the few-photon probabilities ${\cal P}_0$ (green), $1-{\cal P}_1$ (blue), and ${\cal P}_2$ (brown) for a setup with $N=2$ superconducting qubits, varying different parameters around the set \changeByMing{A of} Table~\ref{parameters}.

In Fig.~\ref{fig:conditions}(a) we \jj{plot} the effect of cavity occupation $|\alpha_{\rm ss}|^2$ (controlled via the cavity drive $\Omega_0$) on the single-photon efficiency of the SPS. \jj{All errors reduce \changeByMing{or at least stay nearly constant when increasing the} cavity population $|\alpha_{\rm ss}|^2$, because a larger cavity-mediated driving $\Omega_{\rm cm}=|\alpha_{\rm ss}|g$ increases the speed at which the qubit is excited and reduces the probability of emitting two consecutive photons. However, the driving strength is limited by our need to have an off-resonant cavity and by} the anharmonicity of the qubits [cf.~Sec.~\ref{sec:parameter_sets}].

Fig.~\ref{fig:conditions}(b) illustrates \jj{the need to find an optimum value of the coupling $g$, within inequalities in Eq.\ (\ref{parameter_conditions}). Initially, increasing $g$ increases the driving $\Omega_{\rm cm}$ and reduces errors. However, a large coupling strength enhances the cavity-mediated dipole-dipole interactions $J_{\rm cm}\sim g^2$ and the collective decay $\gamma_{\rm cm}\sim g^2$. Both result in the transfer of excitations between qubits, increasing the probability of no emission ${\cal P}_0$ and two-photon emission ${\cal P}_2.$ For the parameters \changeByMing{A} in Table~\ref{parameters} we find an optimal operation point around $g/\kappa= \changeByMing{10^{-1}}$.}

In Fig.~\ref{fig:conditions}(c) we analyze the effect of \jj{the coupling to the antenna coupling $\gamma.$ Once more, there is an optimal point that satisfies\ (\ref{parameter_conditions}), balancing the imperfections due to unwanted re-excitations} ($\Omega_{\rm cm}\gg \gamma$) and cavity-mediated effects ($\gamma\ll J_{\rm cm},\gamma_{\rm cm}$). \changeByMing{The minima lay} around $\gamma/\Omega_{\rm cm}\sim [10^{-2},10^{-1}]$ for the parameter set A in Table~\ref{parameters}.

Fig.~\ref{fig:conditions}(d) \jj{shows an optimal operation of the emitters when they are on resonance with the drive $\delta\sim\delta_{\rm cm}$. At this point, a $\pi$ pulse excites the qubits with high fidelity, and the Lamb shift is compensated, \changeByMing{which is the main qubit's frequency shift in our setup.}}

Finally, in Fig.~\ref{fig:conditions}(e) and Fig.~\ref{fig:conditions}(f), we observe that the dephasing $\gamma_\phi$ and the losses into uncontrolled channels $\gamma_{\rm loss}$ \jj{have a negligible influence on} the few-photon statistics ${\cal P}_n$ as long as they satisfy $\gamma_\phi,\gamma_{\rm loss}\ll\gamma$ [cf.~Eq.~(\ref{parameter_conditions})]. \jj{Nevertheless, when $\gamma_\phi,\gamma_{\rm loss}\sim \gamma$ we see a rapid increase of events where no photon is detected ${\cal P}_0.$ This is due to emitters becoming effectively off-resonant from the drive when $\gamma_\phi\sim \gamma$, or due to a decrease in the collection efficiency in the antenna, when $\gamma_{\rm loss}\sim \gamma.$}

\subsection{Efficiency of synchronized N-photon generation}\label{sec:efficiency_NP}

The quantity ${\cal P}_1$ is the probability of generating one photon on a given antenna irrespective of the photons emitted in the rest of the channels. To study SPS synchronization, however, we are interested in the probability $P_N$ of emitting exactly one photon on each of the $N$ available antennas or channels. This can be expressed using projectors \jj{onto single-photon Fock states $\Lambda_1^j = |1\rangle_j\langle 1|$ on each the channels}
\begin{align}
    P_N = {\rm Tr}\lbrace \Lambda_{1}^1\cdots\Lambda_{1}^N \rho_{\rm ext}(t)\rbrace,\label{PNdefff}
\end{align}
An efficient \tomas{generation of $N$ synchronized and independent single-photons} should satisfy $P_N\approx ({\cal P}_1)^N\leq {\cal P}_1$, meaning that the $N$-photon product state (\ref{nphotonState}) is generated with high fidelity. \jj{This happens for all parameter sets in Table~\ref{parameters}. The two-photon generation probability satisfies $P_2\approx({\cal P}_1)^2\lesssim 1,$ and deviates from unity because of errors in the single-photon efficiency ${\cal P}_1\lesssim 1$.}

To quantify more precisely the \tomas{independence of the $N$ emitted photons}, we \changeByMing{compute} the $N$-photon \tomas{dependence or demultiplexing} error \changeByMing{$D_N$ from the definition in Eq.~(\ref{scalingwithN})} as,
\begin{align}
D_{N} = \frac{P_{N}}{({\cal P}_1)^N}-1,\label{MultiplexingError}
\end{align}
which describes the deviation from the ideal limit of $N$ perfectly independent and synchronized SPSs.

\jj{We have computed $P_N$ and $D_N$ in the simplest case of two SPSs \changeByMing{as an example}, using the same photon counting methods introduced in Appendix~\ref{sec:counting}. The results of these simulations are shown as} red curves of Figs.~\ref{fig:conditions}(a)-(f) [cf.~red curves and right vertical axis]. We observe that $D_2$ is nearly insensitive to changes in the cavity occupation $|\alpha_{\rm ss}|^2$, the detuning $\delta$, the dephasing $\gamma_\phi/\gamma$, and out-coupling efficiency $\gamma_{\rm loss}/\gamma$, with typical values on order $D_2\sim \changeByMing{10^{-4}}$ for parameters around set \changeByMing{A} of Table~\ref{parameters}. Nevertheless, the variables $g$ and $\gamma$ can drastically change $D_2$ in various orders of magnitude. In particular, for larger cavity coupling $g$ or smaller antenna decay $\gamma$, $D_2$ increases because the cavity-mediated correlations $\sim g^2$ become more important in the timescale of the photon emission $\sim 15/\gamma$ [cf.~right inequality in Eq.~(\ref{parameter_conditions})].

Finally, we note that the exact pulse shape $f(t)$ is not relevant for the \tomas{synchronization} dynamics as long as it approximately induces a $\pi$-pulse on the qubits. To study this, we have performed the calculations in Figs.~\ref{fig:conditions}(a)-(f) using an ideal square pulse and a smooth step pulse $f(t) = \frac{1}{2}\tanh(\frac{t}{\tau_r})\{2 - \tanh(\frac{T}{\tau_r}) - \tanh(\frac{t - T}{\tau_r})\}$ with duration $T$ and ramp time $\tau_r$. When optimizing over $T$ and $\tau_r$ of the smooth pulse, we obtain a deviation smaller than $1\%$ with respect to the result of the simpler square $\pi$-pulse with $T=\pi/(2\Omega_{\rm cm})$. Therefore, in the remainder of the paper we safely consider the ideal square pulse, but we keep in mind that a realistic experimental pulse shape gives similar results when properly optimized. 

\begin{figure*}[ht!]
\includegraphics[width=\textwidth]{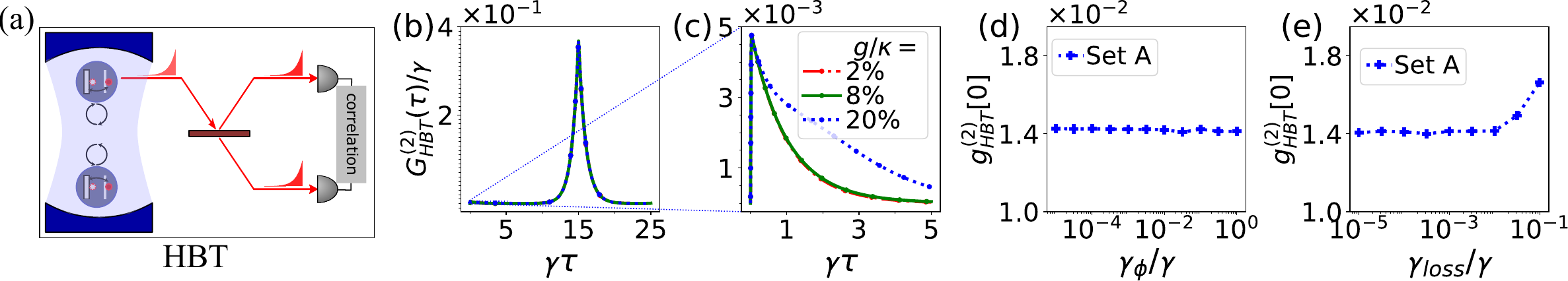}
\caption{Multi-photon contamination of the SPS measured via the second-order correlation function. (a) Schematic of the Hanbury Brown and Twiss (HBT) experiment using a beam-splitter (BS) and two-photon coincidence measurements. (b) HBT correlation function $G^{(2)}_{\rm HBT}(\tau)$ as a function of delay $\tau$, in the case of a train of \changeByMing{2} excitation pulses with repetition rate $R=\gamma/15$. Parameters correspond to set \changeByMing{A} of Table~\ref{parameters}. (c) Enlarged plot of $G^{(2)}_{\rm HBT}(\tau)$ at the zeroth peak, $\tau\sim 0$, for three values of couplings $g/\kappa=\changeByMing{[2,8,20]\cdot 10^{-2}}$. (d,e) Normalized correlation function $g^{(2)}_{\rm HBT}[0]$ at zero delay, as function of dephasing $\gamma_{\phi}/\gamma$ and out-coupling inefficiency $\gamma_{\rm loss}/\gamma$, respectively. Decoherence has a small detrimental effect on the observed multi-photon contamination via $g^{(2)}_{\rm HBT}[0]$. \label{HBT_interference}}
\end{figure*}

\subsection{Single-photon purity}\label{sec:purity}

A standard figure of merit to experimentally quantify the amount of multi-photon contamination of a SPS is the second-order correlation function $G_{\rm HBT}^{(2)}$ \cite{kiraz_quantum-dot_2004,fischer_dynamical_2016}. As \jj{sketched} in Fig.~\ref{HBT_interference}(a), this is measured in a Hanbury Brown and Twiss (HBT) setup, where the output of a given SPS is beam-splitted and measured via coincidences in two intensity detectors. For pulsed emission, the correlation function $G_{\rm HBT}^{(2)}(\tau)$ is defined as \cite{kiraz_quantum-dot_2004,fischer_dynamical_2016} 
\begin{align}
G_{\rm HBT}^{(2)}(\tau) \!=\!& \int\!\mathrm{d}t\,\langle b^{j\dag}_{\rm out}(t)b^{j\dag}_{\rm out}(t+\tau)b^{j}_{\rm out}(t+\tau)b^{j}_{\rm out}(t)\rangle,\label{G2HBT}
\end{align}
where $\tau$ is the time delay between the two-photon detections, and $b^{j}_{\rm out}(t)$ is the annihilation operator \jj{for an outgoing photon on the $j$-th antenna\ \cite{QuantumNoise}. Appendix~\ref{sec:correlations} contains} details on the input-output theory and the calculation of these correlations. For simplicity, our analysis assumes a homogeneous setup and thus we omit the $j$ index in $G_{\rm HBT}^{(2)}(\tau)$. 

In Fig.~\ref{HBT_interference}(b) we show the behaviour of the correlation $G_{\rm HBT}^{(2)}(\tau)$ in our setup \changeByMing{for parameters A} and a train of \changeByMing{2 consecutive} excitation pulses with a repetition rate 
\begin{align}
    R=\gamma/15.
\end{align}
We observe a clear peak at the repetition time \changeByMing{$\tau=1/R$}, corresponding to the detection of two photons coming from different pulses. Relevant to characterize the few-photon statistics of the SPS is the small peak that appears near zero time delay, $\tau\sim 0.$ \jj{A small area in this peak signals a small probability of two- and multi-photon emission}\ \cite{kiraz_quantum-dot_2004}. In Fig.~\ref{HBT_interference}(c) we enlarge the region $\tau\sim 0$ and show that the area of the zeroth peak increases with $g/\kappa,$ \jj{consistently with the behavior} of ${\cal P}_2$ in Fig.~\ref{fig:conditions}(b) for the same parameters.

To quantify more precisely the amount of multi-photon contamination of the SPS, and in a way that is independent of the input power, it is convenient to define the normalized correlation $g_{\rm HBT}^{(2)}[0]$ at zero delay as \cite{fischer_dynamical_2016} 
\begin{align}
    g_{\rm HBT}^{(2)}[0]=\frac{\int_0^{1/(2R)} d\tau G_{\rm HBT}^{(2)}(\tau)}{\left(\int_0^{1/(2R)}dt \langle b_{\rm out}^{j\dag}(t)b_{\rm out}^{j}(t)\rangle\right)^2}.\label{g2HBTdef1}
\end{align}
Here, the numerator corresponds to the area of the zeroth peak at $\tau\sim 0$ and the normalization is the area of the high peak at $\tau=1/R$, and thus $g_{\rm HBT}^{(2)}[0]\sim {\cal P}_2/({\cal P}_1)^2$. In Figs.~\ref{HBT_interference}(d)-(e), we plot the normalized correlation $g_{\rm HBT}^{(2)}[0]$ as a function of $\gamma_{\phi}/\gamma$ and $\gamma_{\rm loss}/\gamma$, respectively. We observe that $g_{\rm HBT}^{(2)}[0]$ is very insensitive to both types of decoherence, which is also consistent with the behavior of ${\cal P}_2$ in Fig.~\ref{fig:conditions}(e)-(f) for the same parameters. \changeByMing{For parameters A,} we obtain a two-photon contamination of $g_{\rm HBT}^{(2)}[0]\changeByMing{\sim 1.4\cdot 10^{-2}}$. \changeByMing{We also calculated this quantity for parameters B-D of Table~\ref{parameters}, for which we obtained $g_{\rm HBT}^{(2)}[0]\sim 2.8\cdot 10^{-2}$, $g_{\rm HBT}^{(2)}[0]\sim 4.4\cdot 10^{-3}$, and $g_{\rm HBT}^{(2)}[0]\sim 1.3\cdot 10^{-2}$, respectively.} Finally, we note that the single-photon purity $p$ is typically defined as $p=1-g_{\rm HBT}^{(2)}[0]$, \changeByMing{which for parameter sets A, B, C, and D read $p= 0.986$, $p=0.972$, $p=0.996$, and $p=0.987$, respectively. We see that slower sources have a larger single-photon purity (see A and C compared to B and D).} 

\subsection{Indistinguishability}\label{sec:indistinguishability}

\begin{figure*}[ht!]
\center
\includegraphics[width=1\textwidth]{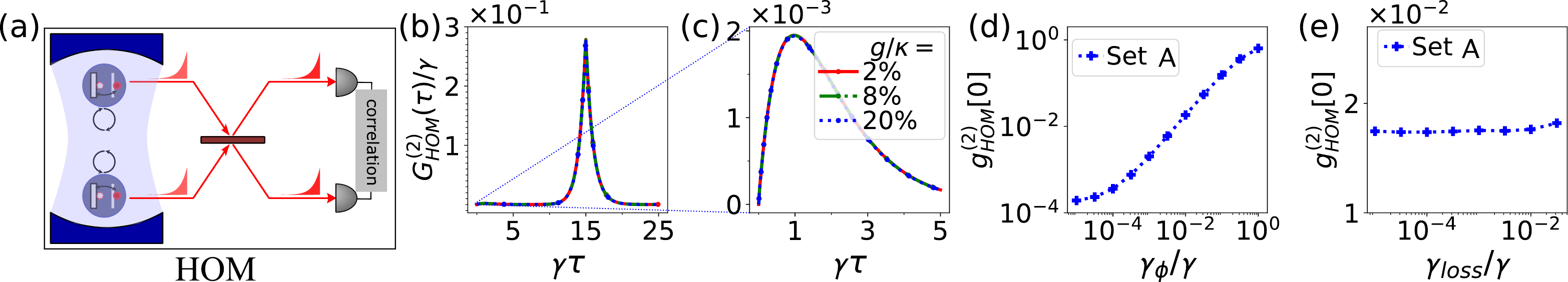}
\caption{Indistinguishability of single-photons from two different SPSs quantified by Hong-Ou-Mandel (HOM) interference. (a) Schematic of the HOM setup using a beam splitter to interfere two nearly independent photons, followed by two-photon coincidence measurements. (b) HOM correlation function $G^{(2)}_{\rm HOM}(\tau)$ as a function of delay $\tau$, in the case of a train of \changeByMing{2} excitation pulses with repetition rate $R=\gamma/15$. Parameters correspond to set \changeByMing{A} of Table~\ref{parameters}. (c) Enlarged plot of $G^{(2)}_{\rm HOM}(\tau)$ at the zeroth peak, $\tau\sim 0$, for three values of couplings $g/\kappa=\changeByMing{[2,8,20]\cdot 10^{-2}}$. (d,e) Normalized correlation function $g^{(2)}_{\rm HOM}[0]$ at zero delay, as function of dephasing $\gamma_{\phi}/\gamma$ and out-coupling inefficiency $\gamma_{\rm loss}/\gamma$, respectively. The indistinguishability dramatically reduces with dephasing, a behavior in good agreement with the analytical prediction $g^{(2)}_{\rm HOM}[0] \approx (\changeByMing{2}\gamma_\phi/\gamma)/(1+\changeByMing{2}\gamma_\phi/\gamma)$ \cite{bylander_interference_2003}.
\label{HOM_interference}}
\end{figure*}

Another important aspect of a high-performance \tomas{multiphoton demultiplexing} scheme is the degree of indistinguishability of the photons emitted by two different sources \cite{senellart_high-performance_2017}. Experimentally, the distinguishability of two \jj{sources} is typically quantified using Hong-Ou-Mandel (HOM) interference as shown in Fig.~\ref{HOM_interference}(a). \jj{Photons coming from two sources interfere on a beam-splitter (BS), and a pair of detectors count coincidences.} The result of these measurements are the HOM correlations $G_{\rm HOM}^{(2)}(\tau)$, which \jj{for pulsed emission} read \cite{fischer_dynamical_2016} 
\begin{align}
G_{\rm HOM}^{(2)}(\tau) \!=\! \int dt \langle b_\text{BS}^{1\dag}(t)b_{BS}^{2\dag}(t+\tau)b_{BS}^2(t+\tau)b_{BS}^1(t)\rangle,
\end{align}
where $\tau$ is the delay time between the two-photon detections. In addition, $b_{BS}^{1}(t)=[b^{1}_{\rm out}(t)+ b^{2}_{\rm out}(t)]/\sqrt{2}$, and $b_{BS}^{2}(t)=[b^{1}_{\rm out}(t)- b^{2}_{\rm out}(t)]/\sqrt{2}$, describe the output fields of photons after interfering at the BS [cf.~Appendix~\ref{sec:correlations}].

Due to the bosonic nature of the emitted photons, two perfectly indistinguishable photons will bunch on either output port of the BS, resulting in a vanishing HOM correlation $G_{\rm HOM}^{(2)}(\tau)$ at zero time delay $\tau\sim 0$. This is perfect HOM interference, and any deviation from it (assuming an ideal BS) can be used to quantify the distinguishability of the generated single-photons.

\jj{A standard figure of merit of} indistinguishability \jj{that accounts for losses and other imperfections} is the normalized HOM correlation function $g_{\rm HOM}^{(2)}[0]$ \cite{fischer_dynamical_2016} 
\begin{align}
    g_{\rm HOM}^{(2)}[0]=\frac{\int_0^{1/(2R)} d\tau G_{\rm HOM}^{(2)}(\tau)}{\prod_{k=1}^2\left(\int_0^{1/(2R)}dt \langle b_{BS}^{k\dag}(t)b_{BS}^{k}(t)\rangle\right)}.\label{g2HOMdef1}
\end{align}
In analogy to Eq.~(\ref{g2HBTdef1}), the numerator of Eq.~(\ref{g2HOMdef1}) corresponds to the area below the zeroth peak, and the normalization to the area of the high peak at the repetition time $\tau=\changeByMing{1/R}$. \jj{The indistinguishability of two SPSs can be simply defined as $I=1-g^{(2)}_{\rm HOM}[0].$}

\jj{We have computed $G_{\rm HOM}^{(2)}(\tau)$ as a function of $\tau$, for a setup with two emitters, \changeByMing{parameters A of Table~\ref{parameters}}, and a train of \changeByMing{2} excitation pulses} [cf.~Fig.~\ref{HOM_interference}(b)]. Here, we observe a clear peak at the repetition time $\changeByMing{\tau=1/R}$ due to the detection of two nearly independent photons coming from two different pulses. The correlations $G_{\rm HOM}^{(2)}(\tau)$ are strongly suppressed at the origin, $\tau\sim 0$, showing only a minor zeroth peak which is enlarged in Fig.~\ref{HOM_interference}(c). The small area of this zeroth peak manifests the high indistinguishability of the single-photons. We also see that this behavior is nearly insensitive to the coupling $g/\kappa$, but may strongly depend on the decoherence parameters. 

Figures\ \ref{HOM_interference}(d)-(e) display $g^{(2)}_{\rm HOM}[0]$ as a function of the decoherence rates $\gamma_\phi/\gamma$ and $\gamma_{\rm loss}/\gamma.$ \jj{Note that in our implementation with \changeByMing{parameters A we reach $I=0.986$, whereas for parameters B-D we obtain $I=0.993$, $I=0.998$, and $I=0.992$, respectively (cf.\ Table~\ref{parameters})}. We also find} that the distinguishability $g^{(2)}_{\rm HOM}[0]$ is quite insensitive to $\gamma_{\rm loss}/\gamma$, but it dramatically increases with dephasing, \jj{following} the analytical prediction $g^{(2)}_{\rm HOM}[0] \approx (\changeByMing{2}\gamma_\phi/\gamma)/(1+\changeByMing{2}\gamma_\phi/\gamma)$ \cite{bylander_interference_2003}. \jj{This contrasts with the behaviour of $D_2$ in Fig.~\ref{fig:conditions}(e), illustrating that the dependence error $D_2$ of two SPSs is qualitatively different from the distinguishability $g^{(2)}_{\rm HOM}[0]$ measured by HOM interference. Consequently, both figures of merit have to be considered when designing} high-performance multiplexed and synchronized SPSs. 

\section{Scalability of synchronized multiphoton generation}\label{sec:scaling}

\begin{figure*}[ht!]
\center
\includegraphics[width=1\textwidth]{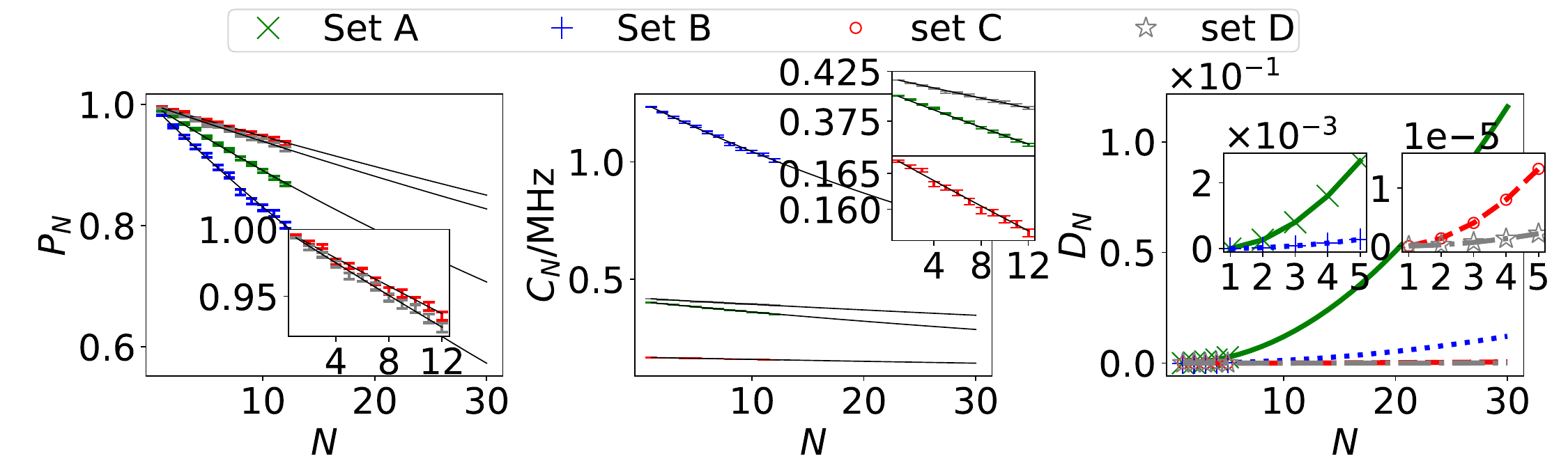}
\caption{Scalability performance of the \tomas{synchronized multiphoton generation} scheme. (a) $N$-photon generation efficiency $P_N$ as a function of $N$, for parameter sets A (\changeByMing{green}), B (\changeByMing{blue}), C (\changeByMing{red}), and D (grey), as given in Table~\ref{parameters}. Data points up to $N=12$ correspond to QT calculations averaged over $M=6000$ trajectories, which show excellent agreement (within the statistical error $\Delta P_N\sim 10^{-3}$) with the prediction for perfectly independent and synchronized SPSs, i.e.~$P_N\approx ({\cal P}_1)^N$. [cf.~black lines and inset]. (b) $N$-photon generation rate $C_N=RP_N$ for pulsed excitation with repetition rate $R=\gamma/15$. Despite parameter set \changeByMing{B} shows the lowest efficiency $P_N$, the faster repetition rate $R\approx 1.26$MHz allows for appreciably higher $C_N$. (c) \tomas{Demultiplexing} error $D_N$ up to $N=5$, and quadratic fit $D_N = \epsilon N (N-1)$ (cf.~solid lines and inset). The fitted scaling factor depends on the system parameters and reads \changeByMing{$\epsilon_A = 1.3\times 10^{-4}$, $\epsilon_B = 1.2\times 10^{-5}$, $\epsilon_C = 6.7\times 10^{-7}$, and $\epsilon_D = 1.0\times 10^{-7}$} for parameter sets A, B, C, and D, respectively. The extrapolation of all these calculations is valid as long as \changeByMing{$D_N\ll 1$}.}
\label{fig:scaling}
\end{figure*}

We are now in a position to discuss the most important results of the present work: the scalability performance of the \tomas{multiphoton emitter}. To do so, we analyze the $N$-photon generation probability $P_N$ and the \tomas{demultiplexing} error $D_N$ as a function of $N$, \changeByMing{for the four parameter sets of Table\ \ref{parameters}.}

We calculate $P_N$ in Eq.~(\ref{PNdefff}) by \jj{solving the master equation} (\ref{master_D_AE}), which includes all qubit-qubit correlations induced by the cavity ($\sim J_{\rm cm}, \gamma_{\rm cm}$), as well as the decoherence effects ($\sim \gamma_\phi,\gamma_{\rm loss}$). The Hilbert space of the system grows exponentially as $2^N$, but \jj{using quantum trajectories (QT) [cf.~Appendix~\ref{sec:counting_QT}] we can estimate} the photon statistics of the \tomas{multiphoton} emission and thereby $P_N$ up to moderately large numbers of SPSs. Figure.~\ref{fig:scaling}(a) displays $P_N$ as a function of $N$ computed from an average over $M=6000$ trajectories and considering the \changeByMing{four} parameter sets of Table~\ref{parameters}. \changeByMing{In all these} cases and up to $N=12$, we numerically confirm that the SPSs are nearly perfectly synchronized and independent, satisfying $P_N\approx({\cal P}_1)^N$ within the statistical error $\Delta P_N\sim 10^{-3}$ [cf.~black lines in Fig.~(\ref{fig:scaling})]. The reduction of $N$-photon generation probability with $N$ is thus mainly limited by the imperfections in the single-photon efficiency ${\cal P}_1$ and not by the \tomas{synchronizing and demultiplexing} scheme. \changeByMing{For the state-of-the-art parameters A and B, we predict a $10$-photon efficiency of $P_{10}\approx 0.90$ and $0.83$, respectively [cf.~inset in Fig.~\ref{fig:scaling}(a)]. The reduction of efficiency for B is mainly due to the larger $\gamma$, but this also allows it to have a much larger $10$-photon emission rate of $C_{10}= 1.0$ MHz compared to $C_{10}=0.4$ MHz for A [cf.~Fig.~\ref{fig:scaling}(b)]. The long-term parameters C and D present similar $N$-photon emission rates, but much larger efficiencies of $P_{10}\approx 0.94$ and $0.91$, respectively}.

To quantify more precisely the deviations from the ideal scaling $P_N\approx ({\cal P}_1)^N$ in Figs.~\ref{fig:scaling}(a)-(b), we compute the \tomas{$N$-photon dependence or demultiplexing} error $D_N$ as defined in Eq.~(\ref{MultiplexingError}). For $N=2$ we know this error is as small as $D_2\sim \changeByMing{10^{-4}}$ [cf.~Fig.~\ref{fig:conditions}] and therefore the QT calculations with an uncertainty $\Delta P_N\sim 10^{-3}$ do not provide enough precision. \jj{To address this problem, we developed a} photon counting approach based on the master equation [cf.~Appendix \ref{sec:counting_ME}], which does not suffer from any statistical uncertainty. In this alternative method, we simulate the photon counters at each antenna by an additional two-level system. \jj{This increases the Hilbert space dimension to $2^{2N}$ and thus limits the numerical computations to a maximum of $N\sim 5$ emitters.} In the inset of Fig.~\ref{fig:scaling}(c) we show the results of $D_N$ as a function $N$ for the four parameter sets of Table~\ref{parameters}, and up to $N=5$. Importantly, we observe that the data is very well approximated by the quadratic fit \changeByMing{$D_N = \epsilon N(N-1)$ stated in Eq.~(\ref{DNapprox}), with a} scaling factor $\epsilon\ll 1$ that depends on the system parameters. \changeByMing{For parameter sets A, B, C, and D, we obtain $\epsilon_A = 1.3\times 10^{-4}$, $\epsilon_B = 1.2\times 10^{-5}$, $\epsilon_C = 6.7\times 10^{-7}$, and $\epsilon_D = 1.0\times 10^{-7}$, respectively  [cf.~inset of Fig.~\ref{fig:scaling}(c)]. Notice that $\epsilon$ is smaller when the conditions $\gamma_{\rm cm}, J_{\rm cm}\ll \gamma$ are better fulfilled overall, confirming that the synchronization performance is limited mainly by residual cavity-induced correlations [cf.~Eq.~(\ref{parameter_conditions})].}

\begin{figure}[t!]
\center
\includegraphics[width=0.8\columnwidth]{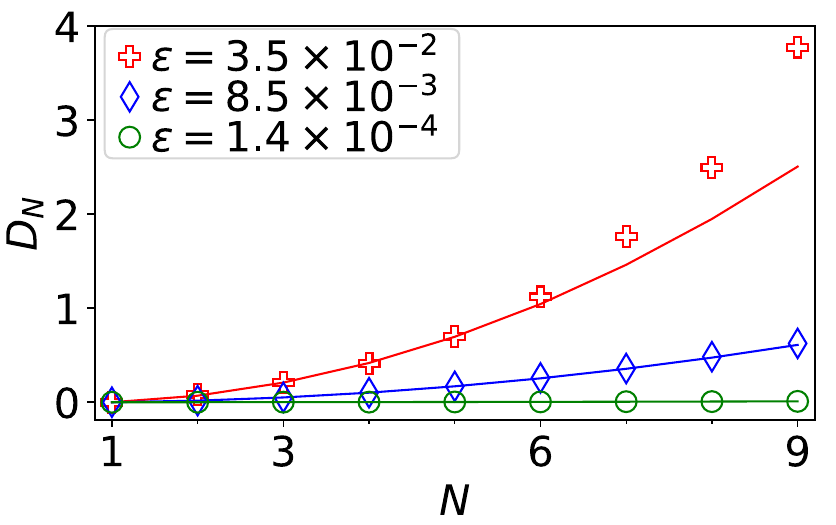}
\caption{Demultiplexing error $D_N$ calculated up to $N=9$ with QT averaged over $M=10^4$ trajectories. \changeByMing{Green circles correspond to the parameter set A of Table\ \ref{parameters}, whereas the other two data sets only differ in having $\gamma/2\pi=0.1$MHz (blue diamonds) and $[\gamma/2\pi,g/2\pi]=[0.1,6]$MHz (red crosses).} Solid lines correspond to fits of $D_N=\epsilon N(N-1)$ up to $N=5$, which result in the $\epsilon$ factors shown in the legend. We observe that the quadratic scaling deviates only for \changeByMing{$D_N\gtrsim 1$}, confirming the validity range \changeByMing{$D_N\ll 1$}.}
\label{fig:limits}
\end{figure}

\tomram{We will now analyze the validity limits of the quadratic scaling $D_N=\epsilon N(N-1)$. This is hard to do for the optimized parameter A-D because the deviation is expected for $N\gg 10$ due to the low correlation errors $\epsilon$. However, we can reduce $\gamma$ and/or increase $g$ to observe the growth of the demultiplexing error with $N$ up to $D_N\gtrsim 1$ with moderate system sizes $N\lesssim 10$.} In Fig.~\ref{fig:limits}, we illustrate this analysis by calculating $D_N$ for \tomram{two modified parameter sets similar to A but with a 10-fold lower antenna decay $\gamma/2\pi=0.1$ MHz (blue diamonds), and with that lower decay and also larger coupling $[\gamma/2\pi,g/2\pi]=[0.1,6]$MHz (red crosses). We also include the results for the standard parameter set A as a reference (blue circles).} We perform quadratic fits $D_N=\epsilon N(N-1)$ to the three data sets up to $N=5$, obtaining \tomram{increasing correlation errors: $\epsilon=1.4\times 10^{-4}$ (green), $\epsilon=8.5\times 10^{-3}$ (blue), and $\epsilon=3.5\times 10^{-2}$ (red). Remarkably, we indeed observe} the on-set of deviations from \tomram{the quadratic} scaling only when $D_N\gtrsim 1$, which occurs for the red data at $N\sim 7$ [cf.~solid red line in Fig.~\ref{fig:limits}]. For $D_N\ll 1$, \changeByMing{all these} data sets show excellent agreement with the quadratic scaling $D_N=\epsilon N(N-1)$, and this constitutes the range of validity of this approximation.

After analyzing the \tomas{demultiplexing} error $D_N$ as a function of $N$, and identifying the limits of scalability in the condition $D_N\ll 1$, we can safely extrapolate the results in Figs.~\ref{fig:scaling}(a)-(c) up to a large $N\gg 1$, only limited \tomram{by the correlation error} as $N\ll \epsilon^{-1/2}$. \tomram{As a criterion for the limit, we use $D_{N_{\rm max}}\sim 0.1$, where $N_{\rm max}$ is the maximum number of sources that can be synchronized without appreciable correlations, i.e.~$P_N\sim ({\cal P}_1)^N$}. Evaluating for the scaling factors obtained for parameters A-D (in the range $\epsilon\sim \changeByMing{10^{-7}-10^{-4}}$) [cf.~Fig.~\ref{fig:scaling}(c) and caption], we \tomram{predict} that our \tomas{multiphoton generation} scheme can be scaled up \changeByMing{to $N_{\rm max}^A\sim 30$ for parameter A, $N_{\rm max}^B\sim 100$ for B, $N_{\rm max}^C\sim 400$ for C, and even reach $N_{\rm max}^D\sim 1000$ for D.} In particular, for state-of-art parameters A, we predict a 30-photon generation probability of $P_{30}\sim(0.989)^{30}\sim 0.72$, with a high generation rate of $C_{30}\sim \changeByMing{300}$kHz [cf.~solid black lines in Figs.~\ref{fig:scaling}(a)-(b)], whereas for state-of-art parameters B, one can reach $P_{100}\sim (0.982)^{100}\sim 0.16$ with a 100-photon generation rate of $C_{100}\sim \tomram{200}$kHz. This is orders of magnitude more efficient than the $3$- to $14$-photon generation and coincidence rates in the range $\sim$kHz-mHz that \tomram{have} been reported in \tomram{single-photon} boson sampling experiments \cite{lenzini_active_2017,wang_high-efficiency_2017,loredo_boson_2017,wang_boson_2019,anton_interfacing_2019,hummel_efficient_2019}.

\section{Inhomogeneity effects}\label{sec:disorder}

So far, we have analyzed the performance of the scheme assuming a perfectly homogeneous setup. In large-scale implementations, however, the SPS parameters will unavoidably present some degree of inhomogeneity. In this section, we analyze the impact of these imperfections.

We consider inhomogeneity deviations on all qubit parameters, which we generically denote by
\begin{align}
    y_j = y + \delta y_j,\label{devs}
\end{align}
where $y=\omega_q, g, \gamma, \gamma_\phi, \gamma_{\rm loss}$ are the average qubit parameters discussed in the previous sections [cf.~Table~\ref{parameters}] and $\delta y_j$ are random deviations over each of them. To statistically describe each of these deviations $\delta y_j$, we assume that they are distributed according to a Gaussian probability distribution, 
\begin{align}
    \Pi(\delta y_j)=\frac{1}{\sqrt{2\pi (\Delta y)^2}}{\rm exp}\left(-\frac{1}{2}\left[\frac{\delta y_j}{\Delta y}\right]^2\right).
\end{align}
Here, $\Delta y$ denotes the standard deviation associated with the disorder on each qubit parameter $y=\omega_q, g, \gamma, \gamma_\phi, \gamma_{\rm loss}$. To quantify the impact of each of the inhomogeneities $\delta y_j$ on the \tomas{performance of the multiphoton emitter}, we compute the average $N$-photon efficiency $\langle\!\langle P_N \rangle\!\rangle_y$ as
\begin{align}
    \langle\!\langle P_N \rangle\!\rangle_y = \frac{1}{M}\sum_{m=1}^M P_N[m]_y,
\end{align}
where the probabilities $P_N[m]_y$ are calculated with the master equation method of Appendix~\ref{sec:counting_ME} for each realization $m=1,\dots,M$ of the independent disorder $y$. 

\begin{figure}[t!]
\center
\includegraphics[width=0.95\columnwidth]{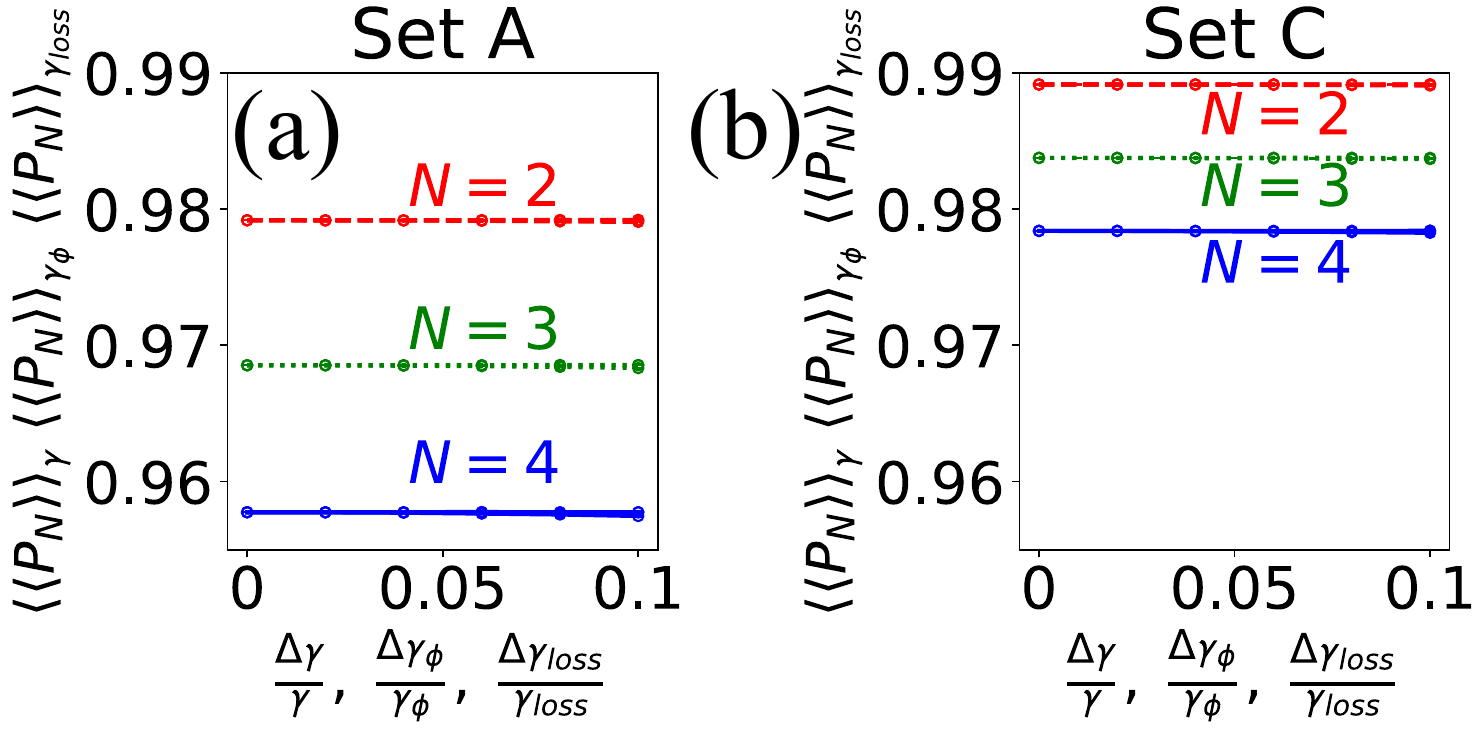}
\caption{Average $N$-photon generation probabilities $\langle\!\langle P_N\rangle\!\rangle_y$, as a function of disorder in antenna decay $y= \Delta\gamma/\gamma$, qubit dephasing $y=\Delta\gamma_{\phi}/\gamma_\phi$, and waveguide loss $y=\Delta\gamma_{\rm loss}/\gamma_{\rm loss}$. (a)-(b) Up to a disorder strength of $10\%$, we do not observe any impact of these three disorder sources on the synchronization and \tomas{demultiplexing} dynamics, as calculated up to $N=4$ for parameter sets \changeByMing{A} and C, respectively. All results are obtained from averaging $M=10^3$ realizations for each type of disorder.} 
\label{fig:inhomo_decoherence}
\end{figure}

The results of these computations for $\langle\!\langle P_N \rangle\!\rangle_y$ in the case of disorder on the antenna couplings $\Delta\gamma$, as well as in the decoherence rates $\Delta\gamma_\phi$ and $\Delta\gamma_{\rm loss}$, are shown in Fig.~(\ref{fig:inhomo_decoherence}). We perform calculations up to $N=4$  slightly inhomogeneous photon sources, for parameters \changeByMing{A} and C, and we do not observe any detrimental effect up to a disorder strength of $10\%$ of the average values [cf.~Fig.~\ref{fig:inhomo_decoherence}(a)-(b)]. This is not surprising in the case of inhomogeneous decoherence rates $\gamma_\phi^j$ and $\gamma_{\rm loss}^j$ since their effects have nothing to do with the synchronization dynamics and are very marginally small anyways $\gamma_\phi^j,\gamma_{\rm loss}^j\ll \gamma$. In the case of inhomogeneous antenna couplings $\gamma_j=\gamma+\delta\gamma_j$, they imply slightly different emission time scales for the emitted single-photons, 
\begin{align}
t\gg \frac{1}{\gamma+\delta\gamma_j}=\frac{1}{\gamma}\left(1-\frac{\delta\gamma_j}{\gamma}+{\cal O}^2\right),
\end{align}
but since we consider a long average waiting time $t= 15/\gamma$, the effect of $10\%$ inhomogeneity in $\delta\gamma_j/\gamma$, is negligible on that timescale.

On the other hand, inhomogeneities in the qubit frequencies $\delta\omega^j_q$ and in the couplings $\delta g_j$ are much more harmful to the \tomas{multiphoton} synchronization performance and therefore they need to be controlled more precisely in an experimental implementation of the device. Inhomogeneous couplings $g_j$, in particular, induce different cavity-mediated driving strengths $\Omega^j_{\rm cm}=\Omega_{\rm cm}+|\alpha_{\rm ss}|\delta g_j$, and therefore different times to realize an exact $\pi$-pulse on each qubit $\sim \pi/(2\Omega_{\rm cm}^j)$. Since we control only the global duration of the cavity pulse, we optimally set it to the average $\pi$-pulse time $T = \pi/(2\Omega_{\rm cm})$, but this unavoidably leads to slightly different probabilities $P_e^j$ of preparing the excited states $|e\rangle_j$ on each qubit. Explicitly, we have
\begin{align}
P_{e}^j \approx \sin^2\left(\frac{\pi}{2}\frac{\Omega_{\rm cm}^j}{\Omega_{\rm cm}}\right)= 1 - \frac{\pi^2}{4}\left(\frac{\delta g_{j}}{g}\right)^2+{\cal O}^4,
\end{align}
and therefore a low coupling disorder $\delta g_j\ll g$ is required for a high performance of the synchronized multiphoton emitter. 

\begin{figure}[t!]
\center
\includegraphics[width=1\columnwidth]{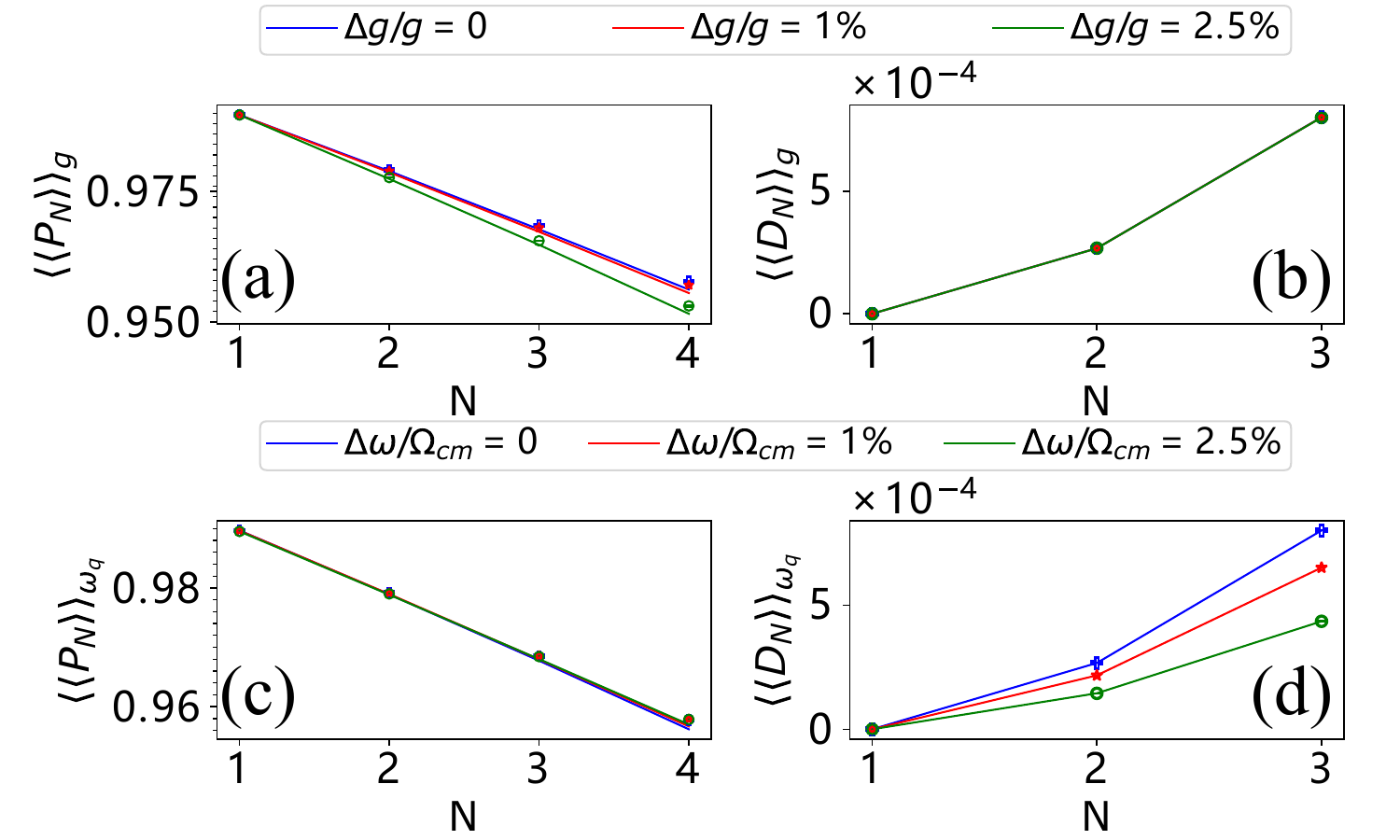}
\caption{Impact of inhomogeneities in qubit-cavity couplings $\delta g_j$ and qubit frequencies $\delta\omega_j$ on the performance of the \tomas{synchronized multiphoton emitter}. (a) Average $N$-photon generation probabilities $\langle\!\langle P_N\rangle\!\rangle_{g}$ up to $N=4$ for coupling disorder strengths $\Delta g/g=0$ (blue), $\Delta g/g=\changeByMing{1}\%$ (red), and $\Delta g/g=\changeByMing{2.5}\%$ (green). (b) Average \tomas{demultiplexing} error $\langle\!\langle D_N \rangle\!\rangle_g$ up to $N=3$, for the same disorder values as in (a). (c)-(d) Analogous calculations as in (a)-(b), but considering $\langle\!\langle P_N\rangle\!\rangle_{\omega_q}$ and $\langle\!\langle D_N\rangle\!\rangle_{\omega_q}$ with disorder in the qubit frequencies $\delta\omega^j_q$. We consider disorder strengths $\Delta\omega_q/\Omega_{\rm cm}=0$ (blue), $\Delta\omega_q/\Omega_{\rm cm}=1\%$ (red), and $\Delta\omega_q/\Omega_{\rm cm}=2.5\%$ (green). All calculations are performed using the parameter set \changeByMing{A} of Table~\ref{parameters} and up to \changeByMing{$M=1.7\times 10^5$} realizations of each type of disorder.}
\label{fig:inhomogeneities}
\end{figure}

To precisely quantify the deviation from the ideal photon independence condition, we define the average \tomas{dependence or demultiplexing} error $\langle\!\langle D_N \rangle\!\rangle_y$ over disorder $y$ as
\begin{align}
     \langle\!\langle D_N \rangle\!\rangle_y = \frac{\langle\!\langle P_N \rangle\!\rangle_y}{\left(\langle\!\langle {\cal P}_1 \rangle\!\rangle_y\right)^N}-1.
\end{align}
Here, the average single-photon efficiency reads,
\begin{align}
    \langle\!\langle {\cal P}_1 \rangle\!\rangle_y = \frac{1}{MN}\sum_{j=1}^N\sum_{m=1}^M {\cal P}_1^j[m]_y,
\end{align}
with ${\cal P}_1^j[m]_y$ calculated from $m=1,\dots, M$ realizations of the disorder $y$ and for each of the $j=1,\dots,N$ SPSs in the setup. In Figs.~\ref{fig:inhomogeneities}(a)-(b) we compute $\langle\!\langle P_N \rangle\!\rangle_g$ and $\langle\!\langle D_N \rangle\!\rangle_g$ up to $N=4$ for parameter set \changeByMing{A}, and disorder strengths $\Delta g/g=0\%$ (blue), $\Delta g/g=\changeByMing{1}\%$ (red), and $\Delta g/g= \changeByMing{2.5}\%$ (green). To achieve reasonable statistical errors $\Delta D_N\lesssim 10^{-\changeByMing{5}}$, these calculations require $\changeByMing{M\sim 1.7\times 10^5}$ realizations. We observe that the $N$-photon efficiency and dependence, $\langle\!\langle P_N \rangle\!\rangle_g$ and $\langle\!\langle D_N \rangle\!\rangle_g$, are minimally modified \tomram{for} $\delta g_j/g\lesssim \changeByMing{2.5\%}$. \tomram{For the average coupling of $g/2\pi=4$ MHz, considered in parameter sets A and B, suppressing these disorder effects thus requires coupling inhomogeneities in the range $\delta g_j/2\pi\sim 0.1$ MHz. This level of inhomogeneity has already been reported in superconducting circuits \cite{wang_controllable_2020} and may be further reduced by using symmetric ring cavity configurations \cite{huang_superconducting_2021, hazra_ring-resonator-based_2021}. For parameter sets C and D, however, reaching the same level of precision requires a coupling inhomogeneity down to $\delta g_j/2\pi\sim 5 $ kHz which has not been achieved with current superconducting technology, but may be possible in the future by reducing residual stray capacitances in the circuit and also considering symmetric cavity configurations.} 

Finally, we analyze the impact of inhomogeneous qubit frequencies $\omega_q^j=\omega_q+\delta\omega^j_q$, which lead to slightly different qubit detunings with respect to the cavity driving frequency $\omega_d$, i.e. $\delta_j=\delta-\delta\omega_q^j$, where $\delta=\omega_d-\omega_q$ is the average detuning. We can set $\omega_d$ to compensate for the cavity-induced Lamb-shift $\delta=\delta_{\rm cm}$, but the remaining inhomogeneities will lead to slightly different probabilities of preparing the $|e\rangle_j$ states on each qubit, i.e.~ 
\begin{align}
P_e^j \approx 1 -\frac{1}{4}\left(\frac{\delta\omega^j_q}{\Omega_{\rm cm}}\right)^2+ {\cal O}\jj{\left(\frac{\delta\omega^j_q}{\Omega_{\rm cm}}\right)^4.}
\end{align}
Therefore, frequency inhomogeneities should also be controlled $\delta\omega_q^j\ll \Omega_{\rm cm}$, to have a high-quality synchronization and demultiplexing. We discussed in Sec.~\ref{sec:parameter_sets} that the qubit frequencies can be fine-tuned by sending specific DC currents on each antenna channel. With this we can make all qubits nearly resonant up to a tuning imprecision in the range $\Delta\omega_q/2\pi\lesssim 1$MHz \cite{dicarlo_demonstration_2009,barends_coherent_2013,pechal_microwave-controlled_2014,arrangoiz-arriola_resolving_2019}. This means that a frequency disorder of order $\Delta\omega_q/\Omega_{\rm cm}\lesssim 2.5\%$ is achievable with state-of-the-art technology. In Figs.~\ref{fig:inhomogeneities}(c)-(d), we calculate $\langle\!\langle P_N\rangle\!\rangle_{\omega_q}$ and $\langle\!\langle D_N\rangle\!\rangle_{\omega_q}$ up to $N=4$ and for frequency disorder strengths $\Delta\omega_q/\Omega_{\rm cm}= 0\%$ (blue), $\Delta\omega_q/\Omega_{\rm cm}= 1\%$ (red), and $\Delta\omega_q/\Omega_{\rm cm}= 2.5\%$ (green). These results confirm that under these realistic disorder conditions the $N$-photon generation efficiency $\langle\!\langle P_N\rangle\!\rangle_{\omega_q}$ is minimally altered with respect to the homogeneous prediction [cf.~blue data in Fig.~\ref{fig:inhomogeneities}(c)]. Moreover, we observe that the demultiplexing error $\langle\!\langle D_N\rangle\!\rangle_{\omega_q}$ reduces with higher frequency disorder $\Delta\omega_q/\Omega_{\rm cm}$ due to the larger independence of the photon emission processes. This occurs at the expense of reducing the generation efficiency $\langle\!\langle P_N\rangle\!\rangle_{\omega_q}$ so it is not a good limit for our purpose.

\tomram{We conclude that the synchronized multiphoton generation scheme is resilient to moderate disorder in all system \tomram{parameters} and that the required level of homogeneity is achievable in state-of-the-art circuit-QED implementations with parameter sets A and B.}

\changeByMing{\section{Multi-mode nature of the cavity}\label{sec:multi-mode}}

So far we have assumed a mono-mode model for the cavity that synchronizes all emitters. Nevertheless, when increasing the length of the cavity to accommodate more qubits, the free spectral range of the cavity reduces and thus more normal cavity modes can participate in the dynamics \cite{sundaresan_beyond_2015}. In this section, we consider these multi-mode effects and show that they can be described with the same effective model in Eqs.~(\ref{master_D_AE})-(\ref{cm_gamma}), but with a redefinition of the cavity-mediated parameters $\Omega_{\rm cm}$, $\delta_{\rm cm}$, $J_{\rm cm}$, and $\gamma_{\rm cm}$ [cf.~Sec.~\ref{sec:Nature_Multimode}]. We analyze multi-mode effects for various cavity lengths and configurations [cf.~Secs.~(\ref{sec:ModerateSize})-(\ref{sec:RingCavity})] and show how to fine-tune the parameters to recover the same synchronization dynamics discussed previously with a mono-mode model. Moreover, we find that the participation of more cavity modes can even be beneficial for the multi-photon generation scheme as it substantially reduces the required driving strength on the cavity.
\\

\subsection{\changeByMing{Effective model including the multi-mode contributions}}\label{sec:Nature_Multimode}

To account for the multi-mode nature of the cavity, we generalize the Hamiltonian and master equation in Eqs.~(\ref{SystemHamiltonian})-(\ref{MasterEquation}), by considering that the qubits interact with many cavity modes $m$, each of them described by a frequency $\omega_c^m$, a decay $\kappa_j$ and a qubit-cavity coupling $g_{j,m}$ (see App.~\ref{sec:open_dynamics} for more details). These quantities scale with mode number $m\in \mathbb{N}$ as \cite{sundaresan_beyond_2015},
\begin{eqnarray}
\omega_c^m &&= m\omega_c^{1},\label{multi_frequency}\\
\kappa_m &&= m\kappa_1 \label{multi_decay},\\
g_{j,m} &&= \sqrt{m}g_{j,1},\label{multi_coupling}
\end{eqnarray}
with $\omega_c^1$, $\kappa_1$, and $g_{j,1}$ the frequency, decay, and coupling corresponding to the fundamental mode $m=1$. Note that these fundamental parameters in turn depend on the geometry and length $L$ of the cavity, such as $\omega_c^1=\pi c/L$ for an open transmission line resonator, and thus the free spectral range scales as $\sim 1/L$.

It is possible to find parameters for which all cavity modes are in the bad-cavity limit, i.e.~they are off-resonant and weakly coupled to the qubits, $g_{j,m}\ll\kappa_m,|\Delta_m|$, so that their dynamics can be adiabatically eliminated. Here, $\Delta_m = \omega_d - \omega_{c}^m$ is the detuning between cavity modes and the external drive $\omega_d$, which is taken to be nearly resonant to the emitters $\omega_d\approx\omega_j^q$. As shown in Appendices~\ref{sec:displacement}-\ref{sec:adiabatic_elimination}, the effective dynamics of the qubits under the above conditions is indeed described by the same effective Master equation and Hamiltonian in Eqs.~(\ref{master_D_AE})-(\ref{H_D_AE}), but with modified cavity-mediated quantities given by
\begin{align}
\Omega_{\rm cm}^j ={}&\sum_{\omega_c^m\sim\omega_d} \frac{\Omega_0 g_{j,m}}{\sqrt{(\kappa_m/2)^2 + \Delta_m^2}},\label{cm_OmegaMulti}\\
\gamma^{jl}_{\rm cm} ={}& \sum_{\omega_c^m\sim\omega_d}\frac{g_{j, m}g_{l, m}\kappa_m}{ (\kappa_m/2)^2+\Delta_m^2}\label{cm_gammaMulti},\\
J^{jl}_{\rm cm} ={}& \sum_{\omega_c^m\sim\omega_d}\frac{g_{j,m} g_{l,m} \Delta_m}{(\kappa_m/2)^2+\Delta_m^2},\label{cm_deltaMulti}
\end{align}
and $\delta^j_{\rm cm} =J_{\rm cm}^{jj}$.

\begin{figure*}[t!]
\center
\includegraphics[width=0.95\textwidth]{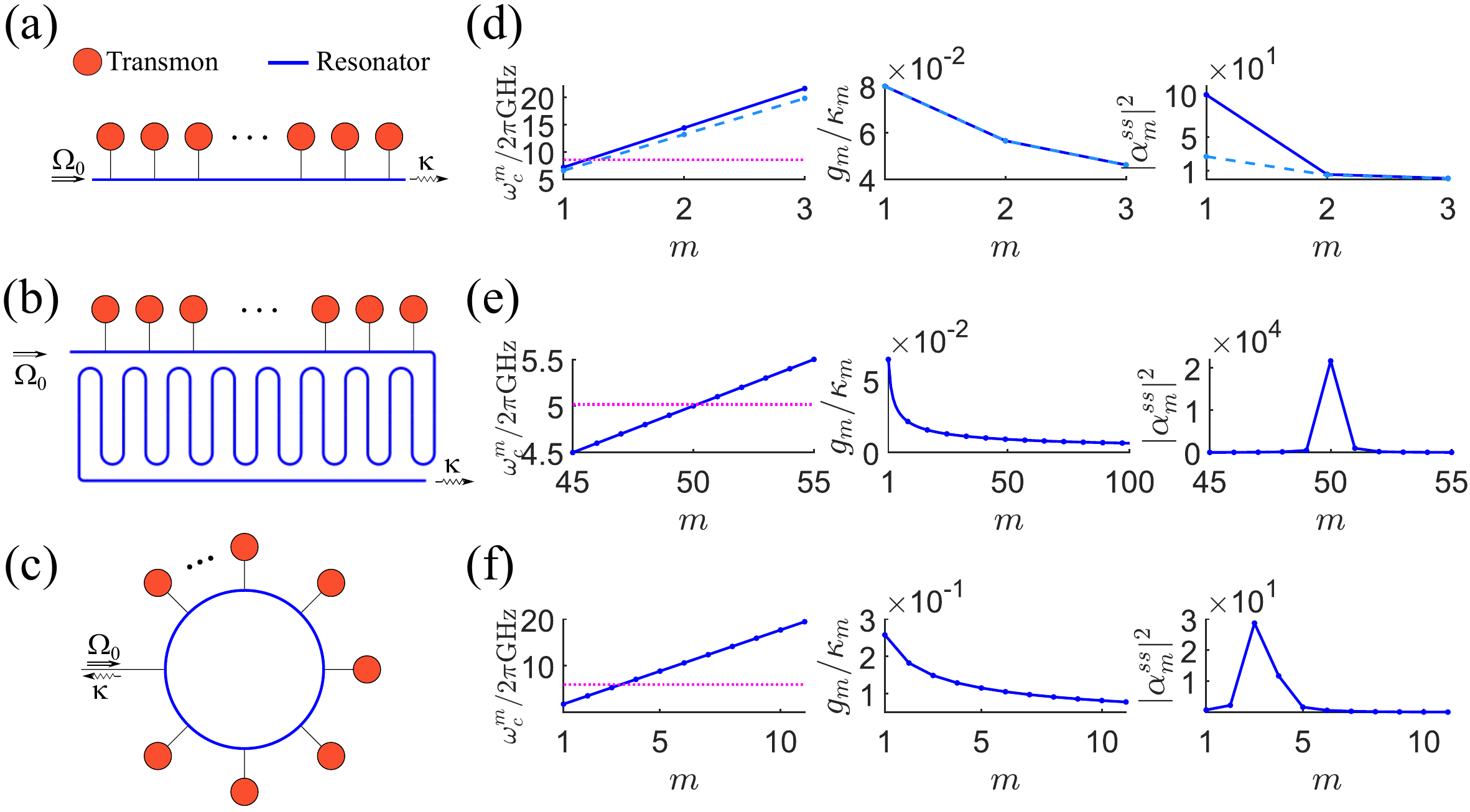}
\caption{Multi-mode effects in the cavity-mediated synchronization of multiple single-photon sources. We consider three different configurations that lead to the same effective dynamics discussed in this work: (a) an open transmission line resonator of moderate length $L=9.17$mm (solid) and $L=8.46$mm (dashed), (b) a long open transmission line resonator with length $L=662$mm, and (c) a ring cavity with length $L=74.6$mm. For each of these configurations (a)-(c), panels (d)-(f) show the mode frequency $\omega_c^m$, the coupling/decay ratio $g_m/\kappa_m$, and the mean steady-state occupation $|\alpha^{\rm ss}_m|^2$ as a function of the mode index $m$. The pink horizontal line in the frequency plots indicates the transition frequency $\omega_q/2\pi$ of the single-photon emitters on each case to see which cavity modes are closer to resonance.}
\label{fig:multimodes}
\end{figure*}

These expressions are the multi-mode generalization of Eqs.~(\ref{cm_Omega})-(\ref{cm_gamma}). They take into account the contribution of all cavity modes, with the largest effect coming from the modes closer to the qubit resonance $\omega_d\sim\omega_j^q$. Therefore, the effect of the multi-mode nature of the cavity is to modify the effective parameters of the dynamics. However, it is always possible to fine-tune the parameters to obtain the same effective physics discussed previously in the manuscript. In the following, we show this on three different multi-mode cavity configurations, namely an open transmission line resonator of moderate length [cf.~Fig.~\ref{fig:multimodes}(a)], a very long transmission line resonator [cf.~Fig.~\ref{fig:multimodes}(b)], and a ring cavity [cf.~Fig.~\ref{fig:multimodes}(c)]. For simplicity, we perform the analysis for homogeneous emitters in the remainder of the section.

\subsection{\changeByMing{Transmission line resonator of moderate size}} \label{sec:ModerateSize}

In this subsection, we show that multi-mode effects are weak in the synchronization setup for cavities of moderate size and when the qubits couple close but still off-resonant to the fundamental resonator mode $m=1$. For this, we consider a transmission line resonator of length $L=9.17$mm [cf.~Fig.~\ref{fig:multimodes}(a)], so that the fundamental frequency reads $\omega_c^1/2\pi= 7.2$ GHz. All other parameters correspond to set A of Table \ref{parameters}, namely $\Delta_1/2\pi = 1400 \textrm{ MHz}$, $g_1/2\pi = 4 \textrm{ MHz}$, $\kappa_1/2\pi = 50 \textrm{ MHz}$, and $\Omega_0/2\pi=14$ GHz. Fig.~\ref{fig:multimodes}(d) displays the mode frequencies $\omega_c^m$, the coupling/decay ratio $g_m/\kappa_m$, and the average occupation $|\alpha^{\rm ss}_m|^2$ as a function of mode label $m$. The pink horizontal line indicates the resonance frequency of the emitters, $\omega_q/2\pi=8.6$ GHz, to compare it with the mode frequencies. We see that all modes satisfy the bad-cavity limit with $g_{m}/\kappa_m<10^{-1}$ and that the occupation is indeed concentrated in the fundamental mode $|\alpha^{\rm ss}_1|^2\sim 100$ as it is closest to resonance. All other modes have occupations more than two orders of magnitude lower and they only slightly modify the effective quantities. Furthermore, if we consider a slightly shorter cavity of length $L=8.46$ mm and fine-tune the system parameters as $\Omega_0/2\pi = 10.4 \textrm{ GHz}$, $\kappa_1/2\pi = 50 \textrm{ MHz}$, and $\Delta_1/2\pi = 2000 \textrm{ MHz}$, we recover the same effective quantities obtained from the mono-mode model in Sec.~\ref{sec:parameter_sets} within a $0.1\%$ error, except for $J_{\rm cm}$ which is even reduced by $28\%$. The modifications of the mode structure and couplings are indicated by the dashed lines in Fig.~\ref{fig:multimodes}(d), where one can see that steady-state occupations are less concentrated in mode $m=1$. The participation of more modes with similar weights allows the system to reduce cavity-mediated couplings as well as the required cavity drive.

\subsection{\changeByMing{Long transmission line resonator}} \label{sec:LongCavity}

In this subsection, we show that the multi-mode structure of the cavity can be exploited to substantially reduce the driving strength required for parameters C and D from $\Omega_0/2\pi\sim 40$ GHz down to $\Omega_0/2\pi\sim 2.64$ GHz, while still inducing the same effective synchronization dynamics on the emitters. The idea is to consider a long resonator so that the free spectral range reduces and more modes can contribute to the effective parameters. In particular, we consider a long open transmission line resonator of length $L=662$mm [cf.~Fig.~\ref{fig:multimodes}(b)], which can fabricated by bending the transmission line inside the chip \cite{sundaresan_beyond_2015,chang_remote_2020}. The length of this resonator is roughly $70$ times larger than in the previous subsection so the mode $m= 50$ of frequency $\omega_c^{50}/2\pi = 5$ GHz becomes the closest to resonance [cf.~left panel of Fig.~\ref{fig:multimodes}(e)]. Here, we consider qubits of frequency $\omega_q/2\pi\approx 4.98$ GHz, and all other parameters are chosen to induce the same effective dynamics as in the mono-mode model with parameters C, namely $g_{50}/2\pi = 0.1 \textrm{ MHz}$, $\kappa_{50}/2\pi = 10.9\textrm{ MHz}$, $\Delta_{50}/2\pi = 17.1\textrm{ MHz}$, and $\Omega_{0}/2\pi = 2.64\textrm{ GHz}$. Note that the driving strength $\Omega_0$ is more than an order of magnitude lower than required in the modo-mode configuration [cf.~Fig.~\ref{fig:multimodes}(a)] This is because of the nearly $10$ off-resonant modes that play a non-negligible role in contributing to the effective parameters [cf.~right panel of Fig.~\ref{fig:multimodes}(e)], and in particular to the cavity-mediated driving $\Omega_{\rm cm}=\sum_m g_m |\alpha_{\rm ss}^m|^2\sim 2\pi\cdot 39$ MHz. Although we have discussed that having a strong resonator drive on order $\Omega_0/2\pi\sim 40$GHz, as required for parameter set C, is not harmful to the superconducting system (as most of this energy is reflected by the cavity), it is instructive to show that the multi-mode structure of the cavity does not introduce detrimental effects and that it may be even beneficial for reducing drive power if needed.

\subsection{\changeByMing{Ring resonator}}\label{sec:RingCavity}

In Sec.~\ref{sec:disorder}, we discussed that the multi-photon synchronizing scheme is resilient to disorder in the couplings $g_j$ as long as this is controlled with a relative precision on the order or below $\sim 2.5\%$. In this subsection, we consider a ring cavity configuration [cf.~Fig.~\ref{fig:multimodes}(c)], which is a very appealing alternative to improve the fabrication of homogeneous couplings $g_{j,m} \approx g_m$ due to its intrinsic circular symmetry. Prototypes of ring cavities \cite{huang_superconducting_2021, hazra_ring-resonator-based_2021} already demonstrate the generation of a very uniform electromagnetic field amplitude, leading to a coupling inhomogeneity below $1 \%$. Note that for the same length $L$ of an open cavity, the ring cavity has a fundamental frequency twice as large $\omega_c^1=2\pi c/L$, due to its periodic boundary conditions. In particular, we consider a ring cavity of length $L = 74.6 \textrm{ mm}$, so that mode $m=3$ with $\omega_c^{3}/2\pi = 5.31$GHz is the closest to the qubit resonance of $\omega_q/2\pi=6$ GHz [cf.~left panel of Fig.~\ref{fig:multimodes}(f)]. Other parameters are taken to induce an effective dynamics identical to parameters A in the mono-mode model, namely $\kappa_3/2\pi = 15.6\textrm{ MHz}$, $g_3/2\pi = 2.32\textrm{ MHz}$, $ \Delta_3/2\pi = 690\textrm{ MHz}$, and $\Omega_{0}/2\pi= 3.7\textrm{ GHz}$. For these parameters, we see that the two modes $m=3,4$ are appreciably populated [cf.~right panel of Fig.~\ref{fig:multimodes}(f)]. As for the open cavity, increasing the length $L$ of the ring cavity also allows accommodating more qubits and reducing the driving strength $\Omega_0$ on the cavity, but the most important added feature of this symmetric configuration is the potential to reduce inhomogeneities in the fabrication of the multi-photon synchronizing device.

\section{Conclusions and Outlook}\label{sec:conclusion}

In this work, we propose a scalable design for \jj{generating efficiently a large number of synchronized, independent, and indistinguishable photons propagating over independent channels}. The \jj{synchronization is provided by a strongly driven resonator in the {\it bad} cavity and weak-coupling limits. The resonator allows for a simultaneous and robust control of all emitters, which deposit photons into individual waveguides, at a high rate and with a high collection efficiency. Although our scheme can be implemented in cavity QED and nanophotonic platforms \cite{welte_photon-mediated_2018,casabone_enhanced_2015,araneda_interference_2018,reitz_coherence_2013,lodahl_interfacing_2015,anton_interfacing_2019,ellis_independent_2018}, we have discussed an efficient} \changeByMing{circuit QED implementation, \cite{pechal_microwave-controlled_2014,zhou_tunable_2020,wang_controllable_2020,barends_coherent_2013,caldwell_parametrically_2018}, where the cavity synchronization mechanism is especially useful to reduce overhead with the number of independent controls \cite{tamate_toward_2022}}.

The only intrinsic limitations for the scalability of the \tomas{synchronized multiphoton device} are the cavity-mediated interactions and collective decay, which can create correlations between SPSs and the emitted photons. Nevertheless, we show that these correlations are strongly suppressed on the timescale of photon emission and that they only induce a nearly negligible quadratic \tomas{demultiplexing} error $D_N\approx \epsilon N(N-1)$ with $\epsilon\sim \tomram{10^{-4}-10^{-5}}$ \changeByMing{for state-of-the-art circuit-QED parameters}. Remarkably, this allows for the synchronization of up to hundreds of nearly independent single-photons, even in the presence of decoherence, disorder, \changeByMing{and multi-mode cavity effects}. 

\jj{Given that, as we show,} each SPS can achieve single-photon efficiency, purity, and indistinguishability above $99\%$, \jj{and the parallel operation in our device enables the efficient creation of large $N$-photon states.} For instance, we predict a $30$-photon probability of $P_{30}\sim 0.72$ at a rate $C_{30}\changeByMing{\sim 300}$kHz and a $100$-photon probability of $P_{100}\sim \changeByMing{0.16}$ at a rate $C_{100}\sim 200$kHz. \jj{This is seven orders of magnitude more efficient than the most sophisticated multiplexed SPSs} up to $N=14$ \changeByMing{single photons} \cite{lenzini_active_2017,wang_high-efficiency_2017,loredo_boson_2017,wang_boson_2019,anton_interfacing_2019,hummel_efficient_2019}. All these promising figures of merit can be further improved when implementing a more refined model for the SPSs such as three-level emitters \cite{fischer_dynamical_2016}, but this is independent of the efficient synchronization scheme we propose.

Scalable and deterministic sources of multi-photon states will be a key resource for realizing large-scale quantum information processing such as quantum optical neural networks \cite{steinbrecher_quantum_2019} or fault-tolerant photonic quantum computation \cite{takeda_toward_2019}. In the short-term, the implementation of our scheme can already enable quantum advantage experiments with hundreds of microwave photons such as boson sampling \cite{brod_photonic_2019} or quantum metrology \cite{motes_linear_2015,ge_distributed_2018}. Moreover, the setup and ideas introduced in this work can be further extended to scale up the generation of correlated multi-photon states \cite{meyer-scott_exponential_2019,pfaff_controlled_2017} with engineered entanglement patterns \cite{besse_realizing_2020}. Many-body methods such as Matrix-product-states \cite{huang_simulating_2019,orus_practical_2014} could be exploited to study the multi-photon correlations of the propagating fields, and recent multi-photon probing methods \cite{ramos_multiphoton_2017,munoz_filtering_2018,dassonneville_number-resolved_2020,lescanne_irreversible_2020,le_jeannic_experimental_2021} could be used to characterize them in the lab.

\section*{Acknowledgements}

The authors thank S.~Paraoanu, F.~Luis, R.~Mart\'inez, R.~Dassonneville, O.~Buisson, Z.~Wang,  E.~Torrontegui, M.~Pino, A.~Gonz\'alez-Tudela, and D.~Porras, for helpful discussions.  Work in Madrid was supported by project PGC2018-094792-B-I00 (MCIU/AEI/FEDER, UE), \changeByMing{Proyecto Sinergico CAM 2020 Y2020/TCS-6545 (NanoQuCo-CM)}, and CSIC Quantum Technology Platform PT-001. T.R. further acknowledges funding from the EU Horizon 2020 program under the Marie Sk\l{}odowska-Curie grant agreement No. 798397 and \changeByMing{the Ram\'on y Cajal program RYC2021-032473-I, financed by MCIN/AEI/10.13039/501100011033 and the European Union NextGenerationEU/PRTR}. M.L. acknowledges support from the China Scholarship Council and the National Natural Science Foundation of China under Grant No. 11475021.

\appendix

\section{Derivation of effective model for cavity synchronization}\label{sec:effective_model}

In this appendix, we outline the derivation of the effective master equation (\ref{master_D_AE}) \changeByMing{starting from general dynamics describing $N$ emitters coupled to a multi-mode cavity [cf.~Sec.~\ref{sec:open_dynamics}].} In particular, we perform a displacement transformation on all the cavity modes [cf.~Sec.~\ref{sec:displacement}], and then an adiabatic elimination of the cavity fluctuations [cf.~Sec.~\ref{sec:adiabatic_elimination}] as shown in the following.

\subsection{\changeByMing{Open quantum dynamics of emitters coupled to a multi-mode cavity}}\label{sec:open_dynamics}

\changeByMing{The master equation and Hamiltonian describing $N$ two-level emitters coupled to a driven multi-mode cavity read,
\begin{align}
\dot{\rho}(t)={}&-i[H(t),\rho]+\sum_m\kappa_m{\cal{D}}[a_m]\rho+\sum_{j=1}^N \gamma_j{\cal{D}}[\sigma_{j}^{-}]\rho\nonumber\\
{}&+\sum_{j=1}^N \gamma_{\rm loss}^j{\cal{D}}[\sigma_{j}^{-}]\rho+\sum_{j=1}^N 2\gamma_\phi^j{\cal{D}},[\sigma_{j}^{+}\sigma_{j}^{-}]\rho,\label{MultimodesMasterEquation}\\
H(t)={}& \sum_{m=1}\omega_c^m a_m^{\dagger}a_m +2\Omega(t)\sum_{m=1}(a_m+a_m^{\dagger})\cos(\omega_d t)\nonumber\\
+{}& \frac{1}{2}\sum_{j=1}^N\omega_j^q\sigma_{j}^{z}+\sum_{j=1}^N\sum_{m=1} g_{j,m}(a_m+a_m^{\dagger})(\sigma_{j}^{-}+\sigma_{j}^{+}).\label{multimodeHam}
\end{align}
Here, the system parameters are the same as in the mono-mode case given in Eqs.~(\ref{SystemHamiltonian})-(\ref{MasterEquation}), except that the cavity is generalized to have many modes $m$, each one described by ladder operators ($a_m$, $a_m^\dag$), a frequency $\omega_c^m$, a decay $\kappa_j$ and a coupling to the qubits $g_{j,m}$.}

\subsection{Coherent displacement of driven cavity modes}\label{sec:displacement}

The quantum Langevin equations \cite{QuantumNoise} associated to the master equation \changeByMing{(\ref{MultimodesMasterEquation})} read
\changeByMing{\begin{align}
\dot{a}_m={}&-\left[\kappa_m/2 + i \omega_c^m\right]a_m - i \sum_{j=1}^N g_{j,m}(\sigma_{j}^{-} + \sigma_{j}^{+})\\
{}&-2i\Omega_m(t)\cos(\omega_d t)-\sqrt{\kappa_m}a_{\rm in}^m,\nonumber\\
\dot{\sigma}_{j}^{-}={}&-\left[    (\gamma_{j}+\gamma^{j}_{\rm loss})/2 +i\omega_j^q\right]\sigma_{j}^{-}\label{QLfirst1}\\
{}&+i\sum_mg_{j,m}(a_m+a_m^{\dag})\sigma_{j}^{z}+\sqrt{\gamma_{j}}\sigma_{j}^{z}b_{\rm in}^{j}\nonumber\\
{}&+\sqrt{\gamma^{j}_{\rm loss}}\sigma_{j}^{z}c_{\rm in}^{j}-i\sqrt{2\gamma_\phi^j}\chi(t)\sigma_j^-,\\
\dot{\sigma}_{j}^{z}={}&-(\gamma_{j}+\gamma^{j}_{\rm loss})(1+\sigma_{j}^{z})\label{QLfirst2}\\
{}&-2i\sum_mg_{j, m}(a_m^{\dag}+a_m)(\sigma_j^{+}-\sigma_j^{-})\nonumber\\
{}&-2\sqrt{\gamma_{j}}\left[\sigma_{j}^{+}b_{\rm in}^{j}
+{\rm h.c.}\right]-2\sqrt{\gamma^{j}_{\rm loss}}\left[\sigma_{j}^{+}c_{\rm in}^{j}
+{\rm h.c.}\right].\nonumber
\end{align}}
Here, $\chi(t)$ is a stochastic white noise dephasing fluctuation satisfying $\langle\!\langle \chi(t)\chi(t')\rangle\!\rangle = \delta(t-t')$ \cite{ramos_correlated_2018}, \changeByMing{$a_{\rm in}^m(t)$ corresponds to the input noise field of the cavity mode $m$}, $b_{\rm in}^{j}(t)$ the input field of photons in antenna $j$, and $c_{\rm in}^{j}(t)$ the photonic input field of unwanted channels coupled to qubit $j$. A coherent driving on the cavity induces a coherent state component on this mode, and thus it is convenient to displace it using the transformation \changeByMing{$a_m\rightarrow \alpha_m+\delta a_m$} as in Eq.~(\ref{cavityDispAlpha}). Here, \changeByMing{$\delta a_m$ corresponds to the quantum fluctuation of the cavity mode around its classical value $\alpha_m$}. By separating the classical and quantum cavity components, the displaced Langevin equation for the \changeByMing{fluctuation of cavity mode $m$} reads,
\changeByMing{\begin{align}
\delta \dot{a}_m={}&-\left[\kappa_m/2 + i \omega_c^m\right]\delta a_m-i\sum_{j=1}^N g_{j, m}(\sigma_{j}^{-}+\sigma_{j}^{+})\nonumber\\
{}&-\sqrt{\kappa_m}a_{\rm in}^m(t),\label{Langevin21}
\end{align}}
and for the qubits,
\changeByMing{\begin{align}
\dot{\sigma}_{j}^{-}={}&-\left[(\gamma_{j}+\gamma^{j}_{\rm loss})/2 +i\omega_j^q\right]\sigma_{j}^{-}+2i\sum_mg_{j, m}{\rm Re}\lbrace \alpha_m \rbrace\sigma_{j}^{z}\nonumber\\
{}&+i\sum_mg_{j, m}(\delta a_m + \delta a_m^{\dag})\sigma_{j}^{z}+\sqrt{\gamma_{j}}\sigma_{j}^{z}b_{\rm in}^{j}\nonumber\\
{}&+\sqrt{\gamma^{j}_{\rm loss}}\sigma_{j}^{z}c_{\rm in}^{j}-i\sqrt{2\gamma_\phi^j}\Theta(t)\sigma_j^-,
\end{align}
\begin{align}
\dot{\sigma}_{j}^{z}={}&-(\gamma_{j}+\gamma_{\rm loss}^{j})(1+\sigma_{j}^{z})-4i\sum_mg_{j, m}{\rm Re}\lbrace \alpha_m \rbrace (\sigma_j^{+}-\sigma_j^{-})\nonumber\\
{}&-2i\sum_mg_{j, m}(\delta a_m^{\dag}+\delta a_m)(\sigma_j^{+}-\sigma_j^{-})\nonumber\\
{}&-2\sqrt{\gamma_{j}}\left[\sigma_{j}^{+}b_{\rm in}^{j}(t)
+{\rm h.c.}\right]\nonumber\\
{}&-2\sqrt{\gamma^j_{\rm loss}}\left[\sigma_{j}^{+}c_{\rm in}^{j}(t)+{\rm h.c.}\right].\label{Langevin23}
\end{align}}

Here, \changeByMing{$\alpha_m(t)$} is determined via the classical differential equation, \changeByMing{ 
\begin{align}
\dot{\alpha}_m(t)={}&-\left[\kappa_m/2 + i\omega_c^m\right]\alpha_m(t)-2i\Omega_0f(t)\cos(\omega_d t),\label{multimodealphaeq}
\end{align}
which is a multi-mode generalization of Eq.~(\ref{alphaSS}).} The master equation associated to the above Langevin equations (\ref{Langevin21})-(\ref{Langevin23}) with the displaced cavity reads
\changeByMing{\begin{align}
\dot{\rho}'(t)={}&-i[H'(t),\rho']+\sum_m\kappa_m{\cal{D}}[\delta a_m]\rho'+\sum_{j=1}^N \gamma_j{\cal{D}}[\sigma_{j}^{-}]\rho'\nonumber\\
{}&+\sum_{j=1}^N \gamma^j_{\rm loss}{\cal{D}}[\sigma_{j}^{-}]\rho'+\sum_{j=1}^N 2\gamma_\phi^j{\cal{D}}[\sigma_{j}^{+}\sigma_{j}^{-}]\rho',\label{master_displaced_noRWA}
\end{align}}
where the system Hamiltonian after the displacement transformation $H'(t)$ is given by,
\changeByMing{\begin{align}
H'(t) ={}& \sum_m \omega_c^m\delta a_m^{\dag}\delta a_m + \sum_{j=1}^N \frac{\omega_{j}^q}{2}\sigma_{j}^{z}\nonumber\\
{}&+\sum_{j=1}^N \sum_mg_{j,m}(\delta a_m + \delta a_m^{\dag})(\sigma_{j}^{-}+\sigma_{j}^{+})\nonumber\\
{}&+\sum_{j=1}^N \sum_m 2g_{j, m}{\rm Re}\lbrace \alpha_m(t) \rbrace(\sigma_{j}^{+}+\sigma_{j}^{-}).\label{Ham_displaced_noRWA}
\end{align}}
Notice that, the displaced master equation (\ref{master_displaced_noRWA}) has the same shape as Eq.~(\ref{MasterEquation}) \changeByMing{except for the inclusion of multi-mode nature and is expressed in terms of the fluctuations $\delta a_m$}. The driving term in displaced Hamiltonian (\ref{Ham_displaced_noRWA}) acts on the qubits rather than on the cavity modes and this is the effective cavity-mediated driving on the qubits that allows for the synchronization of many SPSs in our scheme.

Equation (\ref{master_displaced_noRWA}) is very efficient to perform numerical simulations of the dynamics in the case of strong cavity driving \changeByMing{$|\Omega_0|\gg \kappa_m, |\omega_c^m-\omega_d|$} because most of the cavity photons can be taken into account by the classical field \changeByMing{$\alpha_m(t)$}, and the fluctuations are only weakly populated, \changeByMing{$\langle\delta a_m^{\dagger}\delta a_m\rangle\ll 1$}. \changeByMing{The steady state solution of Eq.~(\ref{multimodealphaeq}) with $f(t)=1$ is given by $\alpha_{\rm ss,m}=-i\Omega_0/(\kappa_m/2-i\Delta_m)$, with $\Delta_m=\omega_d-\omega_c^m$ the detuning of the $m$-th mode with respect to the cavity drive. This implies that only a few cavity modes with $\omega_c^m\sim \omega_d$ have a significant displacement $\alpha_m$ during evolution.} In addition, we consider that all cavity modes are weakly coupled to the emitters, \changeByMing{which are nearly resonant to the cavity drive, i.e.~$g_{j,m}\ll\omega_q^j\approx\omega_d$}, so the dynamics of the system can be further simplified by applying the rotating wave approximation (RWA) to the Hamiltonian in Eq.~(\ref{Ham_displaced_noRWA}). On the one hand, we can approximate the qubit-fluctuation coupling as
\changeByMing{\begin{align}
    \sum_{m}g_{j, m}&(\delta a_m + \delta a_m^{\dag})(\sigma_{j}^{-}+\sigma_{j}^{+})\nonumber\\
    {}&\approx \sum_{\omega_c^m\sim\omega_d}g_{j, m} (\delta a_m^\dag \sigma_j^- +{\rm h.c.}).\label{RWAcoup}
\end{align}}
On the other hand, the cavity-mediated driving on the qubits can also be simplified by applying RWA as
\changeByMing{\begin{align}
    \sum_{m}&g_{j,m}2{\rm Re}\lbrace \alpha_m(t) \rbrace(\sigma_{j}^{+}+\sigma_{j}^{-})\label{RWAdrive}\\
    {}&\approx \sum_{\omega_c^m\sim\omega_d}g_{j, m}|\alpha_{\rm ss, m}|f(t)(\sigma_{j}^{+}e^{-i(\omega_d t-\phi_m)}+{\rm h.c.}),\nonumber
\end{align}
where $|\alpha_{\rm ss, m}|=\Omega_0/\sqrt{(\kappa_m/2)^2+\Delta_m^2}$ and $\phi_m={\rm Arctan}(2\Delta_m/\kappa_m)-\pi/2$} are the steady state amplitude and phase of the cavity displacement, respectively, and $f(t)$ is a step function profile. To demonstrate this last approximation (\ref{RWAdrive}), we first assume the square pulse is switched on $f(t)=1$ and the cavity is initially empty $\alpha_m(0)=0$. Then, the solution of the classical equation (\ref{alphaEq}) for \changeByMing{$\alpha_m(t)$} reads
\changeByMing{\begin{align}
    \alpha_m(t) ={}& |\alpha_{\rm ss, m}|e^{-i(\omega_d t-\phi_m)} (1-e^{-(\kappa_m/2-i\Delta_m)t})\nonumber\\
    {}&-\frac{\Omega_0}{2\omega_d}e^{i\omega_d t}(1-e^{-(\kappa_m/2+2i\omega_d)t}),\label{f1sol}
\end{align}}
where we have also used the inequalities \changeByMing{$|\Delta_m|,\kappa_m\ll \omega_d$} valid for our parameter conditions. The first term in (\ref{f1sol}) reaches the steady state value \changeByMing{$\sim|\alpha_{\rm ss, m}|e^{-i(\omega_d t-\phi_m)}$} in a timescale $\sim 1/\kappa_m$ and the second term causes the fast but low amplitude oscillation observed in Fig.~\ref{fig:dynamics}(a). Even for a large driving strength $\Omega_0\gtrsim \omega_d$, the second term can be neglected with respect to the first when \changeByMing{$|\alpha_{\rm ss, m}|\gg 1$} as it is the case for our parameters. Since in bad cavity limit (\ref{weakcouplingbadcavity}) the timescale for the qubit dynamics is much longer than \changeByMing{$\sim 1/\kappa_m$}, we can use the steady state value \changeByMing{$\alpha_m(t)\approx |\alpha_{\rm ss}|e^{-i(\omega_d t-\phi_m)}$} to approximate the cavity displacement when the drive is switched on $f(t)=1$ as it would be instantaneous for the qubits. Similarly, after the cavity reaches the steady state and it is switched off ($f(t)=0$), the dynamics in Eq.~(\ref{alphaEq}) predicts an exponential decay of \changeByMing{$\alpha_m(t)$} with rate \changeByMing{$\sim \kappa_m$.} In the bad cavity limit we can also approximate this emptying of the cavity as instantaneous for the qubit and thus \changeByMing{$\alpha_m(t)\approx 0$} for $f(t)=0$. \change{In summary, under the above approximations we have \changeByMing{$\alpha_m(t)\approx |\alpha_{\rm ss, m}|e^{-i(\omega_d t-\phi_m)}f(t)$} so that when replacing it in the last term of Eq.~(\ref{Ham_displaced_noRWA}), and applying RWA provided \changeByMing{$g_{j ,m}|\alpha_{\rm ss, m}|\ll \omega_c^m,\omega_q^j\approx\omega_d$}, we obtain Eq.~(\ref{RWAdrive}). Replacing Eqs.~(\ref{RWAcoup})-(\ref{RWAdrive}) in (\ref{Ham_displaced_noRWA}), and going to a rotating frame with respect to the cavity drive frequency $\omega_d$, the dynamics of the system in the displaced picture is finally given by the master equation (\ref{master_displaced_noRWA}) with the RWA Hamiltonian,} 
\changeByMing{\begin{align}
H'(t) \approx{}& - \sum_{\omega_c^m\sim\omega_d}\Delta_m\delta a_m^{\dag}\delta a_m - \frac{1}{2}\sum_{j=1}^N \delta_{j}\sigma_{j}^{z}\nonumber\\
{}&+\sum_{j=1}^N \sum_{\omega_c^m\sim\omega_d}g_{j, m}(\delta a_m^{\dag}\sigma_{j}^{-}+\sigma_{j}^{+}\delta a_m)\nonumber\\
{}&+\sum_{j=1}^N \Omega_{\rm cm}^j(\sigma_{j}^{+}+\sigma_{j}^{-}).\label{Ham_displaced_RWA}
\end{align}}
\changeByMing{Here, $\Omega_{\rm cm}^j=\sum_{\omega_c^m\sim\omega_d}g_{j, m}|\alpha_{\rm ss, m}|$ is the multi-mode generalization of the cavity-mediated driving} in Eq.~(\ref{cm_Omega}) and we have also absorbed the constant phase \changeByMing{$\phi_m$} in the definition of the qubit and fluctuation operators. The qubit detunings read $\delta_j=\omega_d-\omega_q^j$ as in the main text.

\subsection{Adiabatic elimination of cavity fluctuations}\label{sec:adiabatic_elimination}

As discussed in the previous subsection, in the bad cavity limit \changeByMing{$g_{j, m},|\delta_j|\ll \kappa_m$}, the cavity modes reach very quickly a coherent steady state, with its quantum fluctuations $\delta a_m$ close to the vacuum state $|0\rangle$. In this situation, it is convenient to adiabatically eliminate the cavity fluctuation and obtain effective dynamics for the degrees of freedom of the qubits only. To do so, we formally integrate Eq.~(\ref{Langevin21}) and apply Markov approximation provided the qubits' dynamics evolve slowly on the time-scale $\sim 1/\kappa_m$. As a result, we get
\changeByMing{\begin{align}
    \delta a_m (t) \approx{}& -i\sum_j \frac{g_{j,m}}{(\kappa_m/2-i\Delta_m)} \sigma_j^-(t)-\frac{\sqrt{\kappa_m}a_{\rm in}^m(t)}{(\kappa_m/2-i\Delta_m)}\nonumber\\
    {}&-i\sum_j \frac{g_{j,m}}{\kappa_m/2+i(2\omega_d-\Delta_m)}\sigma_j^+(t).
\end{align}}
Replacing the above expressions in the original quantum Langevin equations (\ref{QLfirst1})-(\ref{QLfirst2}), \change{applying RWA, and going to a rotating frame with frequency $\omega_d$}, we obtain the effective dynamics of the qubits, given by the Langevin equations,
\changeByMing{
\begin{align}
\dot{\sigma}_{j}^{-}={}&-\left[(\gamma_{j}+\gamma^{j}_{\rm loss}+\gamma_{\rm cm}^j)/2 -i(\delta_j-\delta^j_{\rm cm})\right]\sigma_{j}^{-}\label{LangevinAE1}\\
{}&+i\Omega_{\rm cm}^jf(t) \sigma_{j}^{z}+\sum_{l\neq j}\sum_{\omega_c^m\sim\omega_d}\frac{g_{j,m}g_{l,m}}{(\kappa_m/2-i\Delta_m)}\sigma_j^z\sigma_l^-\nonumber\\
{}&+\sqrt{\gamma_{j}}\sigma_{j}^{z}b_{\rm in}^{j}+\sqrt{\gamma^{j}_{\rm loss}}\sigma_{j}^{z}c_{\rm in}^{j}-i\sqrt{2\gamma_\phi^j}\Theta(t)\sigma_j^-\nonumber\\{}&+\sum_{\omega_c^m\sim\omega_d}\sqrt{  \frac{(g_{j, m})^{2}\kappa_m}{ (\kappa_m/2)^2+\Delta_m^2}   }\sigma_{j}^{z}a_{\rm in}^m,\nonumber
\end{align}
\begin{align}
\dot{\sigma}_{j}^{z}={}&-(\gamma_{j}+\gamma_{\rm loss}^{j}+\gamma_{\rm cm}^{j})(1+\sigma_{j}^{z})-2i\Omega_{\rm cm}^j f(t)(\sigma_j^{+}-\sigma_j^{-})\nonumber\\
{}&-2\sum_{l\neq j}\sum_{\omega_c^m\sim\omega_d}\left[\frac{g_{j,m}g_{l,m}}{(\kappa_m/2-i\Delta_m)}\sigma_j^+\sigma_l^-+{\rm h.c.}\right]\nonumber\\
{}&-2\sum_{\omega_c^m\sim\omega_d}\sqrt{  \frac{(g_{j, m})^{2}\kappa_m}{ (\kappa_m/2)^2+\Delta_m^2}   }\left[\sigma_{j}^{+}a_{\rm in}^m+{\rm h.c.}\right]\nonumber\\
{}&-2\sqrt{\gamma_{j}}\left[\sigma_{j}^{+}b_{\rm in}^{j}
+{\rm h.c.}\right]-2\sqrt{\gamma^j_{\rm loss}}\left[\sigma_{j}^{+}c_{\rm in}^{j}+{\rm h.c.}\right].\label{LangevinAE2}
\end{align}}

Finally, we can find a master equation whose effective dynamics for the qubits are equivalent to the above quantum Langevin equations (\ref{LangevinAE1})-(\ref{LangevinAE2}). This is the case for the effective master equation (\ref{master_D_AE}) with effective Hamiltonian (\ref{H_D_AE}), \changeByMing{provided the cavity-mediated decay $\gamma_{\rm cm}^{jl}$, detuning, hopping $J_{\rm cm}^{jl}$, driving $\Omega_{\rm cm}^j$, and detuning $\delta_{\rm cm}^j$ are given as in Eqs.~(\ref{cm_OmegaMulti})-(\ref{cm_deltaMulti}), with $\omega_q^j\approx\omega_d$. These expressions reduce to Eqs.~(\ref{cm_Omega})-(\ref{cm_gamma}) in the case of a mono-mode cavity.}

\section{Photon counting and calculation of photon generation probabilities}\label{sec:counting}

Photon detection and counting lies at the heart of Quantum Optics and thus plenty of methods have been developed over time \cite{QuantumNoise,plenio_quantum-jump_1998}. In this Appendix, we describe two methods for quantifying the efficiencies and photon statistics from the output of many SPSs. In Sec.~\ref{sec:counting_ME} we introduce an original photon counting method that we developed based on extending the master equation formalism. Then, in Sec.~\ref{sec:counting_QT}, we discuss a more standard photon counting method using the quantum trajectory (QT) approach. This is useful when the Hilbert space of the system becomes very large but at the expense of losing precision in the computation of the averages. 

\subsection{Photon counting in master equation formalism}\label{sec:counting_ME}

In the master equation dynamics, the information about the emission of photons into the bath is omitted, and therefore one typically resorts to the input-output formalism (cf.~Sec.~\ref{sec:correlations} and Ref.~\cite{QuantumNoise}) to relate measurable photonic quantities to multi-time system correlations. Nevertheless, obtaining the photon statistics from these system correlations involves the computation of multi-dimensional integrals over time, which can be very computationally costly and inefficient when scaling up the number of emitters or photons to probe. 

To tackle the above problem, we developed a non-conventional photon counting method that extends the master equation formalism by reincorporating the information of the emitted photons. To do so, we simulate photon counters at each output channel $j=1,\dots,N$ of the system as quantum ``boxes'' that dynamically count the number of quantum jumps performed by each emitter and thereby the emitted photons. An adequate modeling of the photon counters is crucial to ensure that their presence does not alter the physical dynamics of the system and this is what we detail in the following. First, our method requires extending the Hilbert space of the system ${\cal H}_{\rm sys}$ as
\begin{align}
{\cal H}_{\rm ext}={\cal H}_{\rm sys}\otimes{\cal H}^{\rm c}_1\otimes\dots\otimes{\cal H}^{\rm c}_N, 
\end{align}
where ${\cal H}_j^{\rm c}={\rm span}\lbrace \ket{0}_j,\ket{1}_j,\dots,\ket{N_c}_j\rbrace$ correspond to the extra Hilbert spaces of each photon counter $j=1,\dots,N$ spanned by Fock states $|n\rangle_j$ that count the detected photons from $n=0$ to a maximum value $n=N_c$. The dimension of the extended Hilbert space grows exponentially as 
\begin{align}
{\rm dim}\{{\cal H}_{\rm ext}\}={\rm dim}\{{\cal H}_{\rm sys}\}(1+N_c)^N,
\end{align}
but as long as the total dimension fits in ${\rm dim}\{{\cal H}_{\rm ext}\}\lesssim 2^{10}$ our method provides a fast and efficient way to numerically calculate the few-photon statistics of multiple emitters within a purely master equation approach.  
For any Markovian master equation for a system state $\tilde{\rho}(t)$, one can find an extended master equation that incorporates the few-photon counting statistics in the dynamics. The general recipe is very simple and thus we explain it directly on the \tomas{synchronized multiphoton device} described by Eq.~(\ref{master_D_AE}). In this case, the extended master equation reads
\begin{align}
\dot{\rho}_{\rm ext}={}&-i[\tilde{H}(t),\rho_{\rm ext}]+\sum_{j=1}^N\gamma_j{\cal{D}}[\sigma_{j}^{-}S_{j}^{\dagger}]\rho_{\rm ext}\nonumber\\
{}+&\sum_{j=1}^N\gamma^j_{\rm loss}{\cal{D}}[\sigma_{j}^{-}]\rho_{\rm ext}+\sum_{j=1}^N2\gamma_\phi^j {\cal{D}}[\sigma_{j}^{+}\sigma_{j}^{-}]\rho_{\rm ext}\nonumber\\
{}&+{\cal{D}}\left[\sum_{j=1}^N\sqrt{\gamma^j_{\rm cm}}\sigma_j^-\right]\rho_{\rm ext},\label{mater_3Inhomo}
\end{align}
where $\rho_{\rm ext}$ is the density operator of the extended system including counters. Importantly, the extended master equation looks identical to the original in Eq.~(\ref{master_D_AE}), except for the inclusion of the counting operators $S_j^\dag$ in the Lindblad terms $\sim\gamma_j{\cal{D}}[\sigma_{j}^{-}S_{j}^{\dagger}]$ associated to the photon emissions we want to characterize. The counting operator is a cyclic and unitary operator defined as
\begin{align}
S_j^\dag=\sum_{n=0}^{N_c-1}|n+1\rangle_j\langle n| + |0\rangle_j\langle N_c|,\label{SdagDef}
\end{align}
and therefore every time the emitter $j$ performs a quantum jump and decays into its antenna with rate $\gamma_j$, $S_j^\dag$ adds a new photon to the counter box $j$ as
\begin{align}
S_{j}^{\dagger}|n\rangle_{j} ={}& |n+1\rangle_{j},\quad n \neq N_c.\label{counting1}
\end{align}
When a counter reaches its maximum state $|N_c\rangle_j$, additional quantum jumps would reset the counter as, 
\begin{align}
S_{j}^{\dagger}|N_c\rangle_{j} ={}& |0\rangle_{j},\label{counting2}
\end{align}
and therefore it is very important to choose $N_c$ large enough to avoid reaching this limit and properly account for the physically emitted photons. Using the cyclic definition of $S_j^\dag$ in Eq.~(\ref{SdagDef}), as well as their unitarity properties $S_{j}^{\dagger} S_{j} = S_{j} S_{j}^{\dagger} = 1$, we can take partial trace on the extended master equation (\ref{mater_3Inhomo}) and show that we exactly recover the system dynamics as
\begin{align}
\tilde{\rho}(t) = {\rm Tr}_{\rm c}\{\rho_{\rm ext}(t)\},
\end{align}
where $\tilde{\rho}(t)$ is the state of the system governed by the original master equation without photon counting (\ref{master_D_AE}).

In practical calculations, we therefore solve for the extended state $\rho_{\rm ext}(t)$ in Eq.~(\ref{mater_3Inhomo}), and we then obtain the few-photon statistics of the emitted photons by taking simple expectation values on $\rho_{\rm ext}(t)$. In particular, the probability ${\cal P}_n^j$ to count $n$ photons in channel $j$ is calculated as
\begin{align}
    {\cal P}_n^j(t) = {\rm Tr}\lbrace \Lambda_{n}^{j} \rho_{\rm ext}(t)\rbrace,
\end{align}
where the projection operators $\Lambda_{n}^{j}$ on the counter Fock states $|n\rangle_{j}$ are defined as
\begin{align}
\Lambda_{n}^{j} = |n\rangle_{j} \langle n|.   
\end{align}
More generally, the probability to detect $n_1,\dots,n_N$ photons in output channels $j=1,\dots,N$, respectively, is obtained by products of the projectors as
\begin{align}
    \Pi_{n_1\dots n_N}^{1\dots N}= {\rm Tr}\lbrace \Lambda_{n_1}^{1}\dots\Lambda_{n_N}^{N}\rho_{\rm ext}\rbrace.\label{generalProb}
\end{align}
Using equation (\ref{generalProb}) we can calculate the full few-photon statistics of the system as long as the evolution of the extended master equation (\ref{mater_3Inhomo}) is numerically tractable. This method is particularly suited for characterizing the efficiency of SPSs since in that case we expect the photon statistics to be strongly peaked at $n_j\approx 1$ and therefore taking $N_c=1$ or $N_c=2$ on each counter may be enough. In most calculations shown in the main text, we take $N_c=1$ and quantify the probability of emitting one photon on each of the $N$ channels simultaneously by computing
\begin{align}
    P_N=\Pi_{1\dots 1}^{1\dots N}= {\rm Tr}\lbrace \Lambda_{1}^{1}\dots\Lambda_{1}^{N}\rho_{\rm ext}\rbrace.
\end{align}
In the general case, however, it is important to ensure that the occupation of the last state of the counters is negligible so that the photon statistics are not affected by the finite size of the counters. 

\subsection{Photon counting in quantum trajectories formalism}\label{sec:counting_QT}

The most natural way to implement an ideal photon counting is within the formalism of quantum trajectories (QT) and continuous measurements \cite{plenio_quantum-jump_1998,daley_quantum_2014,dalibard_wave-function_1992,molmer_monte_1993,dum_monte_1992}. Here, the physics of quantum jumps is explicitly simulated during the open system evolution and therefore it is very natural to count them and thereby infer the photon statistics.

The QT interpretation requires re-expressing the master equation (\ref{master_D_AE}) of our SPS \tomas{synchronization and demultiplexing} system as \cite{plenio_quantum-jump_1998,daley_quantum_2014}
\begin{align}
    \dot{\tilde{\rho}}(t)=-i(H_{\rm nh}\tilde{\rho}-\tilde{\rho} H_{\rm nh}^\dag)+\sum_{q=1}^Q c_q \tilde{\rho} c_q^\dag.
\end{align}
Here, $H_{\rm nh}$ is the non-Hermitian Hamiltonian of the system given by
\begin{align}
    H_{\rm nh}= \tilde{H}-\frac{i}{2}\sum_{q=1}^Q c_q^\dag c_q,\label{nhHam}
\end{align}
with $\tilde{H}$ the standard system Hamiltonian in Eq.~(\ref{H_D_AE}) and $c_q$ denoting the $Q=3N+1$ jump operators appearing in the master equation (\ref{master_D_AE}). Using the index $j=1,\dots, N$, which describes each of the $N$ SPSs, the jump operators $c_q$, with $q=1,\dots, Q=3N+1$ can be decomposed as
\begin{align}
    c_{j} = {}& \sqrt{\gamma_j}\sigma_j^-,\label{AntennaJumps}\\
    c_{N+j} = {}& \sqrt{\gamma_{\rm loss}^j}\sigma_j^-,\\
    c_{2N+j} = {}& \sqrt{2\gamma_{\phi}^j}\sigma_j^+\sigma_j^-,\\
    c_{3N+1} = {}& \sum_{j=1}^N\sqrt{\gamma^j_{\rm cm}}\sigma_j^-.
\end{align}

The dynamics of the system in the QT formalism are obtained by calculating the stochastic evolution of $m=1,\dots, M$ realizations of a pure system state $|\Psi_m(t)\rangle$, starting from the initial state $|\Psi_m(0)\rangle=\ket{g}^{\otimes N}$. The evolution of each state realization $|\Psi_m(t)\rangle$ combines deterministic dynamics via the non-Hermitian Schr\"odinger equation
\begin{align}
    \frac{d}{dt}\ket{\Psi_m(t)}=-iH_{\rm nh}\ket{\Psi_m(t)},
\end{align} 
and stochastic quantum jumps that project the quantum state at random times as 
\begin{align}
    |\Psi_m(t)\rangle\rightarrow \frac{c_q|\Psi_m(t)\rangle}{\langle \Psi_m(t)|c_q^\dag c_q |\Psi_m(t)\rangle},
\end{align}
where the specific jump operator $c_q$ is also randomly chosen from the $q=1,\dots, Q$ possibilities on each jump process. When solving for $M\gg 1$ realizations, one can obtain the density matrix $\tilde{\rho}(t)$ of the system from the ensemble average $\tilde{\rho}(t) = (1/M)\sum_{m=1}^M |\Psi_m(t)\rangle \langle\Psi_m(t)|$ or calculate the expectation value of any system operator $X$ as $\langle X(t)\rangle=(1/M)\sum_{m=1}^M\langle \Psi_m(t)| X(t) |\Psi_m(t)\rangle$.

Importantly, if we record the information of how many jumps of each type $c_q$ occurred on each of the $M$ realizations $|\Psi_m(t)\rangle$, we can directly access the photon statistics of the system from this data. For instance, the probability ${\cal P}_n^j$ to generate $n$ photons on antenna $j=1,\dots,N$ is calculated in the QT approach as
\begin{align}
    {\cal P}_n^j = \frac{{\cal N}(c_j|n)}{M},\label{PnQT}
\end{align}
where ${\cal N}(c_j|n)$ denotes the number of trajectories that registered $n$ jumps with a given operator $c_j$ of Eq.~(\ref{AntennaJumps}). Similarly, the $N$-photon probability $P_N$ of generating one photon on each of the $N$ independent antennas can be statistically obtained as
\begin{align}
    P_N = \frac{{\cal N}(c_1,c_2,\dots,c_N|1,1,\dots,1)}{M},\label{PNQT}
\end{align}
where ${\cal N}(c_1,c_2,\dots,c_N|1,1,\dots,1)$ denotes the number of trajectories that registered exactly one jump of each $c_j$ operator in Eq.~(\ref{AntennaJumps}), for $j=1,\dots,N$. 

When calculating many trajectories $M\gg 1$ of the system dynamics, we can gather enough quantum jump data and determine the photon emission probabilities (\ref{PnQT}) and (\ref{PNQT}) with low statistical error. Since this error decreases as $\sim M^{-1/2}$, we typically require on the order of $M\sim 10^{3}-10^4$ trajectories to obtain meaningful results with an error on order $\sim 10^{-2}-10^{-3}$. In practice, this makes the calculation of ${\cal P}^j_n$ and $P_N$ less precise than the extended master equation method in Sec.~\ref{sec:counting_ME}. Nevertheless, the advantage of QT formalism is that we evolve pure states instead of density matrices and that it avoids extending the Hilbert space dimension to include counters as in the extended master equation method. These two key aspects allow us to dramatically reduce the Hilbert space for the simulations and thus to treat much larger systems composed of many more SPSs [cf.~the scalability calculations in Fig.~\ref{fig:scaling}].

\section{Photon correlations and input-output formalism}\label{sec:correlations}

In Secs.~\ref{sec:purity}-\ref{sec:indistinguishability}, we discuss how to quantify multi-photon contamination and photon indistinguishability in the emission of SPSs via second-order photon correlation functions. These photon correlations are measured via coincidence counts either in the Hanbury Brown and Twiss (HBT) or the Hong-Ou-Mandel (HOM) configurations and can be expressed as
\begin{align}
G_{\rm HBT}^{(2),j}(\tau) ={}& \int_{0}^{\infty}\!\!\!\! dt \langle b^{j\dag}_{\rm out}(t)b^{j\dag}_{\rm out}(t+\tau)b^{j}_{\rm out}(t+\tau)b^{j}_{\rm out}(t)\rangle,\label{G2HBTApp}\\
G_{\rm HOM}^{(2),jl}(\tau) \!={}&\!\!\! \int_{0}^{\infty}\!\!\!\!\!\! dt \langle b_{BS}^{1(jl)\dag}\!(t)b_{BS}^{2(jl)\dag}\!(t\!+\!\tau)b_{BS}^{2(jl)}(t\!+\!\tau)b_{BS}^{1(jl)}\!(t)\rangle.\label{G2HOM}
\end{align}
In the HBT correlations (\ref{G2HBTApp}), the operators $b^{j}_{\rm out}(t)$ annihilates an output photon on the antenna channel $j$ at time $t$, and in HOM correlations (\ref{G2HOM}), $b_{BS}^{1(jl)}(t)=[b^{j}_{\rm out}(t)+ b^{l}_{\rm out}(t)]/\sqrt{2}$, and $b_{BS}^{2(jl)}(t)=[b^{j}_{\rm out}(t)- b^{l}_{\rm out}(t)]/\sqrt{2}$ correspond to the photonic output operators after passing through a beamsplitter that connects antennas two antennas $j$ and $k$.

As explained in Secs.~\ref{sec:purity}-\ref{sec:indistinguishability}, it is convenient to define normalized second-order correlation functions at zero time delay, which in the case of pulsed emission read,
\begin{align}
g_{\rm HBT}^{(2),j}[0]{}&=\frac{\int_0^{\frac{1}{2R}} d\tau G_{\rm HBT}^{(2),j}(\tau)}{\left(\int_0^{\frac{1}{2R}}dt \langle b_{\rm out}^{j\dag}(t)b_{\rm out}^{j}(t)\rangle\right)^2},\label{g2HBTnorm}\\
g_{\rm HOM}^{(2),jl}[0]{}&=\frac{\int_0^{\frac{1}{2R}} d\tau G_{\rm HOM}^{(2),jl}(\tau)}{\prod_{k=1}^2\left(\int_0^{\frac{1}{2R}}dt \langle b_{BS}^{k(jl)\dag}(t)b_{BS}^{k(jl)}(t)\rangle\right)}.\label{g2HOMnorm}
\end{align}

To express the photon correlations in Eqs.~(\ref{G2HBTApp})-(\ref{g2HOMnorm}) in terms of two-time system correlations, we can use the input-output relation, which read
\begin{align}
b^{j}_{\rm out}(t) = b^{j}_{\rm in}(t) + \sqrt{\gamma_j} \sigma_j^{-}(t),\label{inputoutputB}
\end{align}
where $\sigma_j^{-}(t)$ is the Pauli operator of qubit $j=1,\dots,N$. In addition, the input field $b^{j}_{\rm in}(t)$ is the same operator that appears in the quantum Langevin equations (\ref{Langevin23}) or (\ref{LangevinAE2}), and it can be expressed as a Fourier transform over the annihilation operators $b_{j}(\omega)$ of photons of frequency $\omega$ propagating in antenna $j$, namely   
\begin{align}
b_{\rm in}^{j}(t) ={}& \frac{1}{\sqrt{2\pi}}\int_{-\infty}^{\infty} d\omega e^{-i\omega t} b_{j}(\omega).
\end{align}

If we use the input-output relation (\ref{inputoutputB}) into the correlations functions in Eqs.~(\ref{g2HBTnorm})-(\ref{g2HOMnorm}), and consider that all antennas are initially in vacuum state $|0\rangle$, we obtain
\begin{align}
g_{\rm HBT}^{(2),j}[0]{}&= \frac{\int_0^{\frac{1}{2R}}\!d\tau\!\int_0^{\infty}\!dt \langle \sigma^{+}_j(t)\sigma^{+}_j(t + \tau)\sigma^{-}_j(t+\tau)\sigma^{-}_j(t)\rangle}{\left(\int_0^{\frac{1}{2R}}dt \langle \sigma_{j}^+(t)\sigma_j^{-}(t)\rangle\right)^2},\label{g2HBTsystem}
\end{align}
and
\begin{align}
g_{\rm HOM}^{(2),jl}[0]{}&=\frac{\int_0^{\frac{1}{2R}}\!d\tau\!\!\int_{0}^{\infty}\!\!dt\langle \xi_{1}^{(jl)\dag}\!(t)\xi_{2}^{(jl)\dag}\!(t\!+\! \tau)\xi_{2}^{(jl)}\!(t\!+\!\tau)\xi^{(jl)}_{1}\!(t)\rangle}{\prod_{k=1}^2\left(\int_0^{\frac{1}{2R}}dt \langle \xi_{k}^{(jl)\dag}(t)\xi_{k}^{(jl)}\!(t)\rangle\right)}.\label{g2HOMsystem}
\end{align}
Here, the superposition system operators between qubits $j$ and $l$ read
\begin{align}
\xi_1^{(jl)}(t)=(\sqrt{\gamma_j}\sigma_j^-(t)+\sqrt{\gamma_l}\sigma_l^-(t))/\sqrt{2},\\ \xi_2^{(jl)}(t)=(\sqrt{\gamma_j}\sigma_j^-(t)-\sqrt{\gamma_l}\sigma_l^-(t))/\sqrt{2}. 
\end{align}

Finally, we can use the Eq.~(\ref{master_D_AE}) together with the quantum fluctuation regression theorem \cite{QuantumNoise} to calculate the system expectation values $\langle \sigma_{j}^+(t)\sigma_j^{-}(t)\rangle$ and $\langle \xi_{k}^{(jl)\dag}(t)\xi_{k}^{(jl)}\!(t)\rangle$, as well as the two-time system correlation functions $\langle \sigma^{+}_j(t)\sigma^{+}_j(t + \tau)\sigma^{-}_j(t+\tau)\sigma^{-}_j(t)\rangle$ and $\langle \xi_{1}^{(jl)\dag}\!(t)\xi_{2}^{(jl)\dag}\!(t\!+\! \tau)\xi_{2}^{(jl)}\!(t\!+\!\tau)\xi^{(jl)}_{1}\!(t)\rangle$. After calculating these quantities, we replace them into Eqs.~(\ref{g2HBTsystem})-(\ref{g2HOMsystem}), and perform the corresponding integrals over time $t$ and time delay $\tau$ to obtain the results for the HBT and HOM second-order correlations functions shown in Secs.~\ref{sec:purity}-\ref{sec:indistinguishability}.

%

\end{document}